\documentclass[11pt]{article}
\usepackage{mathrsfs}
\usepackage[centertags]{amsmath}
\usepackage{amsfonts}
\usepackage{amssymb}
\usepackage{amsthm}
\usepackage{cases}
\usepackage{indentfirst}
\usepackage{epsfig}
\usepackage{graphicx}
\usepackage{color}
\usepackage{amsmath}
\usepackage{floatrow}
\date{}

\newtheorem{theorem}{Theorem}[section]
\newtheorem{corollary}{Corollary}[section]
\newtheorem{lemma}{Lemma}[section]
\newtheorem{remark}{Remark}[section]
\numberwithin{equation}{section}

\def\CTE{\operatorname*{\mathrm{CTE}}}
\def\ECTE{\operatorname*{\mathrm{ECTE}}}

\def\ECTM{\operatorname*{\mathrm{ECTM}}}
\def\RECTM{\operatorname*{\mathrm{RECTM}}}

\def\P{\operatorname*{\mathbb{P}}}
\def\E{\operatorname*{\mathbb{E}}}
\def\R{\operatorname*{\mathbb{R}}}
\def\N{\operatorname*{\mathcal{N}}}
\def\I{\operatorname*{\mathbb{I}}}
\allowdisplaybreaks[4]

\begin{document}
\title{Expectile-based conditional tail moments with covariates}
\author{\small{Qian Xiong\quad Zuoxiang Peng\thanks{Corresponding
author. Email: pzx@swu.edu.cn}}
\\
\\
\small{School of Mathematics and Statistics, Southwest
University, 400715 Chongqing, China}}
\maketitle

\begin{quote}
{\bf Abstract.}~~Expectile, as the minimizer of an asymmetric quadratic loss function, is a coherent risk measure and is helpful to
use more information about the distribution of the considered risk.
In this paper, we propose a new risk measure by replacing quantiles by expectiles,  called
expectile-based conditional tail moment, and focus on
the estimation of this new risk measure as the conditional survival function of the risk, given the risk exceeding the expectile and given a value of the covariates, is heavy tail. Under some regular conditions, asymptotic properties of this new estimator are considered. The extrapolated estimation of the conditional tail moments is also investigated.  These results are illustrated both
on simulated data and on a real insurance data.

{\bf Keywords.}~~Expectiles; Conditional tail moment; Heavy-tailed
distribution; Semi-parametric estimation; Second order regular variation

{\bf AMS 2000 Subject Classification.}~~ Primary 60G70; Secondary 62G32.

\end{quote}

\section{Introduction}\label{sec1}
For a real-valued random variable $Y$ with
a cumulative distribution function (cdf) $F$, Koenker and Bassett (1978) showed
that the quantile at level $\alpha\in(0,1)$ can be obtained
by minimizing asymmetrically weighted mean absolute deviations:
\begin{equation}\label{eq1.1}
  q(\alpha)=\mathop{\arg\min}_{q\in\R} \E(\rho_\alpha(Y-q)-\rho_\alpha(Y)),
\end{equation}
where $\rho_\alpha(y)=\left|\alpha-{\I}_{\{y\leq0\}}\right||y|$
with $\I_{\{\cdot\}}$ the indicator function.
Replacing the absolute deviations in the asymmetric loss
function $\rho_\alpha(y)$ by squared deviations, Newey and Powell (1987)
introduced the concept of $\alpha$-expectile:
\begin{equation}\label{eq1.2}
  e(\alpha)=\mathop{\arg\min}_{q\in\R} \E(\eta_\alpha(Y-q)-\eta_\alpha(Y)),
\end{equation}
where $\eta_\alpha(y)=\left|\alpha-{\I}_{\{y\leq0\}}\right|y^2$.
Obviously, both expectiles and quantiles can be seen as
two prototypical cases of the generalized quantiles,
see e.g., Bellini et al. (2014) and Mao et al. (2015).
It is well known that the quantile plays an important role in risk
management due to its simplicity and interpretability.
However, the quantile is in general not coherent in the sense of Artzner et al. (1999),
and moreover, it is often criticised for only taking into account
the probability of a big loss and not the size of the loss.
Unlike quantiles, expectiles make use of more information on the loss
but its lack of closed form expressions usually makes analysis difficult.
More discussion of expectiles with quantiles, we refer
to Bellini et al. (2014), Bellini and Di Bernardino(2017),
Daouia et al. (2019) and references therein.

Another popular alternative to the quantile and the
expectile is the conditional tail expectation ($\CTE$),
defined by $\CTE(\alpha)=\E[Y|Y>q(\alpha)]$.
Replacing the quantile in the $\CTE$ by the expectile,
another intuitive tail conditional expectation,
named the expectile-based conditional tail expectation ($\ECTE$),
is given by ${\ECTE}(\alpha)=\E[Y|Y>e(\alpha)]$.
This quantity has been used by Taylor (2008) to estimate the $\CTE$
and also introduced in Daouia et al. (2020) recently.
Motivated by this definition, the $\ECTE$ can be generalized to the
expectile-based conditional tail moment ($\ECTM$). It is defined by
\begin{equation*}\label{eq2.1}
  {\ECTM}_k(\alpha)=\E[Y^k|Y>e(\alpha)],
\end{equation*}
where $k\geq0$ is such that $\E[Y^k]<\infty$.
In practice, some auxiliary information may be helpful for
a more precise analysis of the loss function.
To this end, the definition of $\ECTM$ may be extended to the case
where the random variable $Y$ is recorded together with a finite-dimensional
covariate $X\in{\R}^p$. In this situation, we define the regression $\ECTM$ by
\begin{equation}\label{eq2.2}
  {\RECTM}_k(\alpha|x)=\E[Y^k|Y>e(\alpha|x),X=x],
\end{equation}
where $e(\alpha|x)$ is referred to as the conditional
expectile (Girard et al., 2022), given by
\begin{equation}\label{eq2.3}
  e(\alpha|x)=\mathop{\arg\min}_{q\in\R}
  \E(\eta_\alpha(Y-q)-\eta_\alpha(Y)|X=x).
\end{equation}

The objective of this paper is to estimate the ${\RECTM}_k(\alpha|x)$
when the level $\alpha$ is extreme, i.e., $\alpha=\alpha_n\rightarrow1$
as the sample size $n\rightarrow\infty$.
Let $(Y_i,X_i), i=1, \ldots, n$, be independent copies of the
random pair $(Y,X)$. We will denote by $F(\cdot|x)$ and
$\bar{F}(\cdot|x)$ the continuous conditional distribution function and
conditional survival function of $Y$ given $X=x$, respectively.
Also, we will denote by $g$ the probability density function of $X$.
For heavy-tailed distributions, the problem of estimating various risk measures
has been widely studied in the literature. For instance, Daouia et al. (2011)
considered the estimation of conditional quantiles, defined by
$q(\alpha|x)=\inf\{y\in\R: \bar{F}(y|x)\leq1-\alpha\}$.
El Methni et al. (2014, 2018) and Xiong et al. (2023) proposed
non-parametric estimators and semi-parametric estimators
of the regression conditional tail moment, respectively.
Goegebeur et al. (2021b, 2021c) studied the estimation of the conditional
marginal expected shortfall and established their asymptotic properties,
and Goegebeur et al. (2021a) considered an application of the conditional
risk measure to estimate the conditional risk premium in reinsurance.
More recently, Girard et al. (2022) constructed the non-parametric kernel estimator
of the extreme conditional expectile.
Specifically, from Jones (1994), the $e(\alpha|x)$ is precisely
the $\alpha$-quantile of the conditional survival
function $\bar{G}(y|x)$, that is,
\begin{equation*}\label{eq2.4}
  e(\alpha|x)=\inf\{y\in\R: \bar{G}(y|x)\leq1-\alpha\},
\end{equation*}
where
\begin{equation*}\label{eq2.5}
  \bar{G}(y|x)=
  \frac{\psi^{(1)}(y|x)}{2\psi^{(1)}(y|x)+(y-m^{(1)}(x))g(x)}
\end{equation*}
with
\begin{equation*}\label{eq2.6}
  \psi^{(k)}(y|x)\!=\!\E[(Y-y)^k{\I}_{\{Y>y\}}|X=x]g(x)
  ~\text{and}~ m^{(k)}(x)\!=\!\E[Y^k|X=x], \quad\forall k\geq0.
\end{equation*}
Thus, by applying the following kernel estimators:
\begin{equation*}\label{eq2.7}
  \hat{g}_n(x):=\frac{1}{n}\sum_{i=1}^n K_{h_n}(x-X_i),
\end{equation*}
\begin{equation}\label{eq2.8}
  \hat{m}_n^{(1)}(x):=\frac{\frac{1}{n}\sum_{i=1}^n K_{h_n}(x-X_i)Y_i}
  {\frac{1}{n}\sum_{i=1}^n  K_{h_n}(x-X_i)}
\end{equation}
and
\begin{equation*}\label{eq2.9}
  \hat{\psi}_n^{(k)}(y|x):=\frac{1}{n}\sum_{i=1}^n
  K_{h_n}(x-X_i)(Y_i-y)^k{\I}_{\{Y_i>y\}},
\end{equation*}
where $K_{h_n}(\cdot)=K(\cdot/h_n)/h_n^p$ with K a density on $\R^p$ and
$h_n$ is a positive non-random sequence of bandwidths such that
$h_n\rightarrow0$ as $n\rightarrow\infty$, Girard et al. (2022) proposed
the non-parametric estimator for $e(\alpha|x)$ as follows:
\begin{equation}\label{eq2.10}
  \hat{e}_n(\alpha|x):=\inf\{y\in\R: \hat{\bar{G}}_n(y|x)\leq1-\alpha\},
\end{equation}
where
\begin{equation}\label{eq2.11}
  \hat{\bar{G}}_n(y|x):=\frac{\hat{\psi}_n^{(1)}(y|x)}
  {2\hat{\psi}_n^{(1)}(y|x)+(y-\hat{m}_n^{(1)}(x))\hat{g}_n(x)}.
\end{equation}

Throughout the paper, we also restrict ourselves to the class of
conditional heavy-tailed distributions.
In other words, we assume that
$\bar{F}(\cdot|x)$ is regularly varying at infinity with $-1/\gamma(x)<0$,
denoted by $\bar{F}(\cdot|x)\in RV_{-1/\gamma(x)}$, i.e.,
\begin{equation}\label{eq2.12}
  \lim_{y\rightarrow\infty}\frac{\bar{F}(\lambda y|x)}{\bar{F}(y|x)}
  =\lambda^{-1/\gamma(x)}, \quad\forall\lambda>0.
\end{equation}
Moreover, a refined version of condition \eqref{eq2.12} is stated as follows.

\noindent\textbf{Assumption C1.}
Assume that the conditional survival function $\bar{F}(\cdot|x)$ is
continuously differentiable and satisfies the following Von-Mises condition:
\begin{equation*}\label{eq2.19}
  \lim_{y\rightarrow\infty}\frac{y\bar{F}^{'}(y|x)}{\bar{F}(y|x)}=-1/\gamma(x),
  \quad\text{for}\;  \gamma(x)>0.
\end{equation*}
In the remainder of this paper we implicitly assume that
$\gamma(x)<1$ and $\E[Y_{-}|X=x]<\infty$ with $Y_{-}=\max(-Y,0)$.
These two conditions combined imply that $\E[|Y||X=x]<\infty$, and hence
conditional expectiles of $Y$ given $X=x$ are well-defined.
Indeed, under (C1) or equivalently \eqref{eq2.12}, the
conditional quantile $q(1-1/\cdot|x)$ is
regularly varying at infinity with $\gamma(x)>0$, and moreover,
Proposition 1 in Daouia et al. (2020) shows that
$e(\alpha|x)/q(\alpha|x)\rightarrow (1/\gamma(x)-1)^{-\gamma(x)}$
with $\gamma(x)<1$ as $\alpha\rightarrow1$.
Thus, it follows that
\begin{equation}\label{eq2.22}
  \lim_{y\rightarrow\infty}\frac{e(1-1/(\lambda y)|x)}{e(1-1/y|x)}
  =\lambda^{\gamma(x)}, \quad\forall\lambda>0.
\end{equation}

From \eqref{eq2.2}, ${\RECTM}_k(\alpha_n|x)$ can be rewritten as
${\RECTM}_k(\alpha_n|x)
=\frac{\E[Y^k{\I}_{\{Y>e(\alpha_n|x)\}}|X=x]}{\bar{F}(e(\alpha_n|x)|x)}$.
Thus, under assumption (C1), Lemma 1 of EI Methni et al. (2014) gives that
for $k\in[0,1/\gamma(x))$,
\begin{equation}\label{eq2.16}
  {\RECTM}_k(\alpha_n|x)\sim \frac{1}{1-k\gamma(x)}(e(\alpha_n|x))^k
\end{equation}
as $n\rightarrow\infty$. Making use of this asymptotic connection
we obtain the following semi-parametric estimator of ${\RECTM}_k(\alpha_n|x)$:
\begin{equation}\label{eq2.17}
  \widetilde{\RECTM}_{k,n}(\alpha_n|x)
  :=\frac{1}{1-k\widetilde{\gamma_{n}}(x)}(\hat{e}_n(\alpha_n|x))^k,
\end{equation}
where $\hat{e}_n(\alpha_n|x)$ is given by \eqref{eq2.10} and the
bias-reduced estimator of the conditional tail index $\gamma(x)$ is given by
\begin{equation}\label{eq2.51}
  \widetilde{\gamma}_n(x)
  =\widehat{\gamma}_n(x)
  \left(
  1-\frac{\hat{m}^{(1)}(x)\sum_{j=1}^J(\tau_j^{\widehat\gamma_n(x)}-1)}
  {\hat{e}_n(\alpha_n|x)\sum_{j=1}^J\log(1/\tau_j)}
  \right),
\end{equation}
where $\hat{m}^{(1)}(x)$ is given by
\eqref{eq2.8}, and $\widehat{\gamma}(x)$ is, a kernel version of the Hill estimator (Hill 1975) based on the conditional expectiles, given by
\begin{equation}\label{eq2.18}
  \widehat{\gamma}_n(x)=\frac{\sum_{j=1}^J \big(\log\hat{e}_n(1-\tau_j(1-\alpha_n)|x)-\log\hat{e}_n(\alpha_n|x)\big)}
  {\sum_{j=1}^J \log(1/\tau_j)},
\end{equation}
with $(\tau_j)_{j\geq1}$ a positive decreasing sequence of weights.
This estimator $\widehat{\gamma}_n(x)$ can be seen as a
expectile-based analogue of that in Daouia et al. (2011).
More work on estimation of $\gamma(x)$, see Daouia et al. (2013),
Goegebeur et al. (2021c), Girard et al. (2022), and references therein.
Furthermore, in view of \eqref{eq2.16} and \eqref{eq2.22}, it can be found that
assumption (C1) implies that $\RECTM_k(1-1/\cdot|x)\in RV_{k\gamma(x)}$
for $k\in[0,1/\gamma(x))$, and thus, we have ${\RECTM}_k(\beta_n|x)\sim{\RECTM}_k(\alpha_n|x)
\big(\frac{1-\alpha_n}{1-\beta_n}\big)^{k\gamma(x)}$
as $n\rightarrow\infty$.
This asymptotic connection naturally suggests
the Weissman type estimator (Weissman, 1978) as follows
\begin{equation}\label{eq2.43}
  \widetilde{\RECTM}_{k,n}^W(\beta_n|x):=\widetilde{\RECTM}_{k,n}(\alpha_n|x)
  \left(\frac{1-\alpha_n}{1-\beta_n}\right)^{k\widetilde{\gamma}_n(x)},
\end{equation}
where $0<\alpha_n<\beta_n<1$, $\widetilde{\gamma}_n(x)$ and $\widetilde{\RECTM}_{k,n}(\alpha_n|x)$
are given by \eqref{eq2.51} and \eqref{eq2.17}, respectively.
In such a case, $((1-\alpha_n)/(1-\beta_n))^{k{\widetilde\gamma}_n(x)}$ is referred
as the extrapolation factor which is helpful for estimating
${\RECTM}_k(\cdot|x)$ at arbitrarily large levels.

In order to derive the asymptotic properties of the considered estimators,
we conclude this section by introducing some assumptions.

\noindent\textbf{Assumption C2.} Assume that the conditional survival function
$\bar{F}(\cdot|x)$ is second-order regularly varying at infinity with the first-order
parameter $-1/\gamma(x)<0$ and the second-order parameter $\rho(x)\leq0$, denoted by
$\bar{F}(\cdot|x)\in2RV_{-1/\gamma(x), \rho(x)}$, that is, there exists some ultimately
positive or negative function $A(t|x)$ with $A(t|x)\rightarrow0$ as $t\rightarrow\infty$
such that
\begin{equation*}\label{eq2.27}
  \lim_{t\rightarrow\infty} {\frac{1}{A(1/\bar{F}(t|x)|x)}}\left(\frac{\bar{F}(ty|x)}{\bar{F}(t|x)}-y^{-1/\gamma(x)}\right)
  =y^{-1/\gamma(x)}\frac{y^{\rho(x)/\gamma(x)}-1}{\rho(x)\gamma(x)},
  \quad\forall y>0.
\end{equation*}

\noindent\textbf{Assumption C3.} Assume that $m^{(2)}(x)<\infty$
and there exist constants $c>0$ and $r>0$ such that
\begin{equation*}\label{eq2.29}
  |g(x)-g(x')|
  \vee |m^{(1)}(x)-m^{(1)}(x')|
  \vee |m^{(2)}(x)-m^{(2)}(x')|
  \leq c\Vert x-x^{\prime}\Vert, \quad\forall x'\in B(x,r),
\end{equation*}
where $\Vert\cdot\Vert$ denote a norm on $\R^p$, $\vee$ and $B(x,h)$ denote
the maximum operator and the ball centred at $x$ with radius $h>0$, respectively.

\noindent\textbf{Assumption C4.} K is a bounded density function
on $\R^p$, with support $S$ included in the unit ball of $\R^p$ for
the norm $\Vert\cdot\Vert$.

The rest of the paper is organized as follows. Section \ref{sec2} states
the main results for the asymptotic properties of our estimators.
Section \ref{sec3} illustrates the performance of our
estimators on simulated data. An application to a real insurance
data is presented in Section \ref{sec4}. Some auxiliary
lemmas and all proofs are postponed to Section \ref{sec5}.

\section{Main results}\label{sec2}
In this section, we provide the main results.
First, we introduce the notation in Girard et al. (2022):
\begin{equation*}\label{eq2.30}
  \omega_{h_n}(y_n|x)=\sup_{\substack{z\geq y_n\\x'\in B(x,h_n)}}\\
  \frac{1}{\log z}\left|\log\frac{\bar{F}(z|x')}{\bar{F}(z|x)}\right|,
\end{equation*}
and establish the joint central limit theorem for
$\hat{e}_n(\alpha_n|x)$ and $\widehat{\gamma}_n(x)$ in the following theorem.

\begin{theorem}\label{th2.1}
Assume that $(C1)$-$(C4)$ hold with $\gamma<1/2$. Suppose
that there exists $\delta\in(0,1)$ such that $\E[Y_{-}^{2+\delta}|X=x]<\infty$.
Consider a sequence $1=\tau_1>\tau_2>\cdots\tau_J>0$, where $J\geq2$ is a positive integer.
Let $\alpha_n\rightarrow1$,
$h_n\rightarrow0$
be such that
$n{h_n}^p(1-\alpha_n)\rightarrow\infty$
and
$n{h_n}^{p+2}(1-\alpha_n)\rightarrow0$.
If
$\frac{\sqrt{nh_n^p(1-\alpha_n)}}{e(\alpha_n|x)}\rightarrow\lambda(x)\in\R$,
$\sqrt{nh_n^p(1-\alpha_n)}A((1-\alpha_n)^{-1}|x)\rightarrow0$ and
$\sqrt{n{h_n}^p(1-\alpha_n)}\log(1-\alpha_n)
\times\omega_{h_n}((1-\delta)e(\alpha_n|x)|x)\rightarrow0$,
then
\begin{equation*}\label{eq2.32}
  \sqrt{nh_n^p(1-\alpha_n)}
  \left\{
  \left(\frac{\hat{e}_n(\alpha_n|x)}{e(\alpha_n|x)}-1\right),
  \left(\widehat{\gamma}_n(x)-\gamma(x)\right)
  \right\}^T
  \overset{d}{\longrightarrow}
  \N\left(\mu(x),\frac{\Vert K\Vert_2^2\gamma^2(x)}{g(x)}\Lambda(x)\right),
\end{equation*}
where
$\mu(x)=\left(0,\lambda(x)m^{(1)}(x)b(x)\right)^T$ with
$b(x)=\frac{\gamma(x)\sum_{j=1}^J(\tau_j^{\gamma(x)}-1)}
{\sum_{j=1}^J\log(1/\tau_j)}$
and $\Lambda(x)$ is the symmetric matrix having entries
{\footnotesize\begin{eqnarray}
  &&\Lambda_{1,1}(x)=\frac{2\gamma(x)}{1-2\gamma(x)}, \label{eq2.33}\nonumber\\
  &&\Lambda_{1,2}(x)=\Lambda_{2,1}(x)
  =\frac{\sum_{j=2}^J\tau_j^{-\gamma(x)}-J+1}
  {(1-2\gamma(x))\sum_{j=1}^J\log(1/\tau_j)}, \label{eq2.34}\\
  &&\Lambda_{2,2}(x)\!=\!\frac{\frac{2}{1\!-\!2\gamma(x)}
  \!\left(\!(J\!-\!1)^2(1\!-\!\gamma(x))\!+\!\gamma(x)\sum_{j=2}^J\tau_j^{-1}
  \!+\!(1\!-\!J)\sum_{j=2}^J\tau_j^{-\gamma(x)}\!\right)
  \!+\!2{\I}_{\{J>2\}}\sum_{j=2}^{J-1}\sum_{l=j+1}^J\tau_j^{-1}
  \!\left(\frac{(\tau_l/\tau_j)^{-\gamma(x)}}{1-2\gamma(x)}\!-\!1\!\right)}
  {\left(\sum_{j=1}^J\log(1/\tau_j)\right)^2}. \nonumber\\
  &&\label{eq2.35}
\end{eqnarray}}
\end{theorem}

\begin{remark}\label{rem2.1}
Obviously, Theorem \ref{th2.1} gives the asymptotic normality result
for $\hat{e}_n(\alpha_n|x)$ which is also provided in Girard et al. (2022).
Also, we can obtain the asymptotic normality of $\widehat{\gamma}_n(x)$
which shows that $\widehat{\gamma}_n(x)$
is an estimator of $\gamma(x)$ with an asymptotic bias given by
$\frac{\gamma(x)m^{(1)}(x)\sum_{j=1}^J(\tau_j^{\gamma(x)}-1)}
{e(\alpha_n|x)\sum_{j=1}^J\log(1/\tau_j)}$.
To deal with the bias in $\widehat{\gamma}_n(x)$,
we use the estimators for the unknown quantities in the above asymptotic bias and
propose a bias-reduced version of $\widehat{\gamma}_n(x)$ in \eqref{eq2.51}.
\end{remark}

Next, direct computations lead to the following asymptotic normalities.

\begin{corollary}\label{cor2.1}
Under the assumptions of Theorem \ref{th2.1}, we have
\begin{equation*}\label{eq2.52}
  \sqrt{nh_n^p(1-\alpha_n)}
  \left\{
  \left(\frac{\hat{e}_n(\alpha_n|x)}{e(\alpha_n|x)}-1\right),
  \left(\widetilde{\gamma}_n(x)-\gamma(x)\right)
  \right\}^T
  \overset{d}{\longrightarrow}
  \N\left(0,\frac{\Vert K\Vert_2^2\gamma^2(x)}{g(x)}\Lambda(x)\right),
\end{equation*}
where $\Lambda(x)$ is given in Theorem \ref{th2.1}.
\end{corollary}

Now, the joint asymptotic normality of
$\widetilde{\RECTM}_{k,n}(\alpha_n|x)$ and $\hat{e}_n(\alpha_n|x)$
is established in the following statement.

\begin{theorem}\label{th2.2}
Assume that the assumptions of Theorem \ref{th2.1} hold.
If $\sqrt{nh_n^p(1-\alpha_n)}A(1/\bar{F}(e(\alpha_n|x)|x)|x)\to0$ as $n\to\infty$,
then for $\gamma(x)<1/k$, we have
\begin{equation*}\label{eq2.37}
  \sqrt{nh_n^p(1-\alpha_n)}
  \left\{
  \left(\frac{\widetilde{\RECTM}_{k,n}(\alpha_n|x)}{\RECTM_k(\alpha_n|x)}-1\right), \left(\frac{\hat{e}_n(\alpha_n|x)}{e(\alpha_n|x)}-1\right)
  \right\}^T
  \overset{d}{\longrightarrow}
  \N\left(0,\frac{\Vert K\Vert_2^2\gamma^2(x)}{g(x)}V(x)\right),
\end{equation*}
where $V(x)$ is the symmetric matrix having entries
$V_{1,1}(x)=\frac{2k^2\gamma(x)}{1-2\gamma(x)}+\frac{2k^2}{1-k\gamma(x)}\Lambda_{1,2}(x)
+\big(\frac{k}{1-k\gamma(x)}\big)^2\Lambda_{2,2}(x)$,
$V_{1,2}(x)=V_{2,1}(x)=\frac{2k\gamma(x)}{1-2\gamma(x)}+\frac{k}{1-k\gamma(x)}\Lambda_{1,2}(x)$
and $V_{2,2}(x)=\frac{2\gamma(x)}{1-2\gamma(x)}$
with $\Lambda_{1,2}(x)$ and $\Lambda_{2,2}(x)$ given by
\eqref{eq2.34} and \eqref{eq2.35}, respectively.
\end{theorem}

In particular, Theorem \ref{th2.2} provides the following asymptotic
normality of $\widetilde{\RECTM}_{k,n}(\alpha_n|x)$.

\begin{corollary}\label{cor2.2}
Under the assumptions of Theorem \ref{th2.2}, we have for $\gamma(x)<1/k$
\begin{equation*}\label{eq2.41}
  \sqrt{nh_n^p(1-\alpha_n)}
  \left(\frac{\widetilde{\RECTM}_{k,n}(\alpha_n|x)}{\RECTM_k(\alpha_n|x)}-1\right)
  \overset{d}{\longrightarrow}
  \N\left(0,\frac{\Vert K\Vert_2^2\gamma^2(x)}{g(x)}V_{1,1}(x)\right),
\end{equation*}
where $V_{1,1}(x)$ is the one given by Theorem \ref{th2.1}.
\end{corollary}

\begin{remark}\label{rem2.3}
As shown in Theorem \ref{th2.2}, the conditions
$n{h_n}^p(1-\alpha_n)\rightarrow\infty$
and $n{h_n}^{p+2}(1-\alpha_n)\rightarrow0$
restrict the rate of $\alpha_n$ tending to 1.
To overcome this issue, we propose a Weissman
type estimator (Weissman, 1978) in \eqref{eq2.43}
by using an extrapolation tool.
\end{remark}

To end this section, we give the asymptotic distribution of $\widetilde{\RECTM}_{k,n}^W(\beta_n|x)$, which states as follows.

\begin{theorem}\label{th2.3}
Assume that the assumptions of Theorem \ref{th2.2} hold.
Let $\beta_n\rightarrow1$ be such that
$\frac{1-\beta_n}{1-\alpha_n}\rightarrow0$,
and $\frac{\sqrt{nh_n^p(1-\alpha_n)}}{\log((1-\alpha_n)/(1-\beta_n))}\rightarrow\infty$
as $n\rightarrow\infty$.
Then, for $\gamma(x)<1/k$ we have
\begin{equation*}\label{eq2.44}
  \frac{\sqrt{nh_n^p(1-\alpha_n)}}{\log((1-\alpha_n)/(1-\beta_n))}
  \left(\frac{\widetilde{\RECTM}_{k,n}^W(\beta_n|x)}{\RECTM_k(\beta_n|x)}-1\right)
  \overset{d}{\longrightarrow}
  \N\left(0,\frac{k^2\Vert K\Vert_2^2\gamma^2(x)}{g(x)}\Lambda_{2,2}(x)\right),
\end{equation*}
where $\Lambda_{2,2}(x)$ is the one given by \eqref{eq2.35}.
\end{theorem}

\section{Simulation study}\label{sec3}
In this section, the finite sample behavior of our estimators is evaluated
by a simulation study.
Assume that the one-dimensional random covariate $X (p=1)$
is uniformly distributed on the unit interval $[0,1]$.
In our simulation, we set $\gamma(x)=\frac{1}{4}+\frac{\sin(2\pi x)}{20}$
for $0\leq x\leq1$ and consider a Burr distribution with the cdf
$\bar{F}(y|x)=(1+y^{1/\gamma(x)})^{-1}$ for $y>0$.

\begin{figure}[htbp]
  \centering
  \includegraphics[width=6cm]{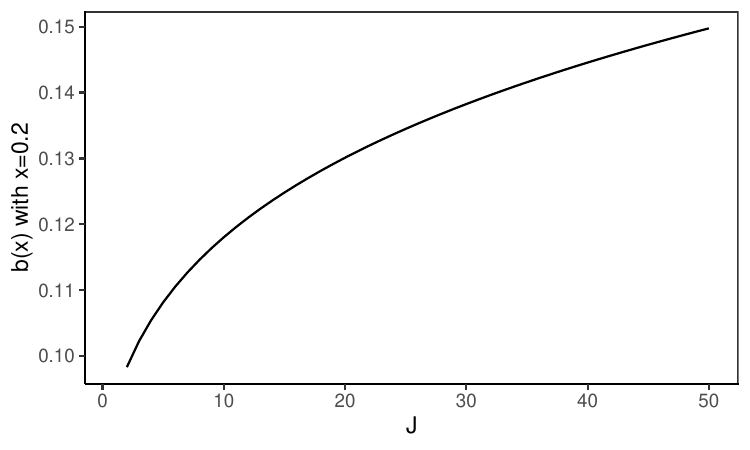}\includegraphics[width=6cm]{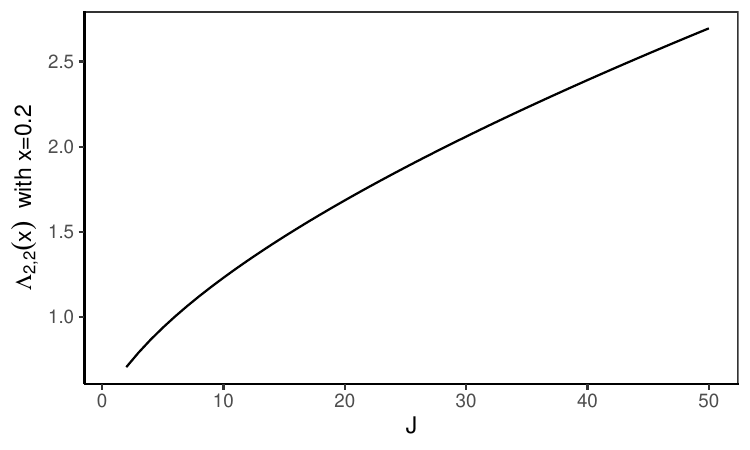}\\
   \includegraphics[width=6cm]{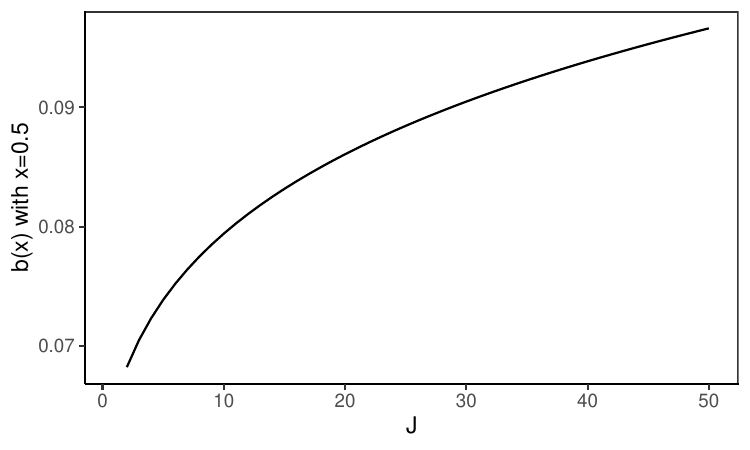}\includegraphics[width=6cm]{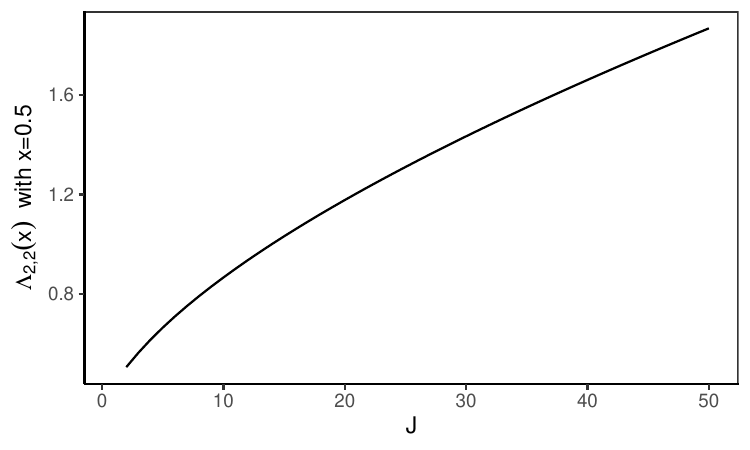}\\
  \caption{Plot of the asymptotic bias term $b(x)$ (left) and the asymptotic variance term $\Lambda_{2,2}(x)$ (right) of the estimator
$\widehat{\gamma}_n(\cdot)$ versus the value of $J$.}
\label{Fig3.1}
\end{figure}

\begin{figure}[htbp]
  \centering
\includegraphics[width=6cm]{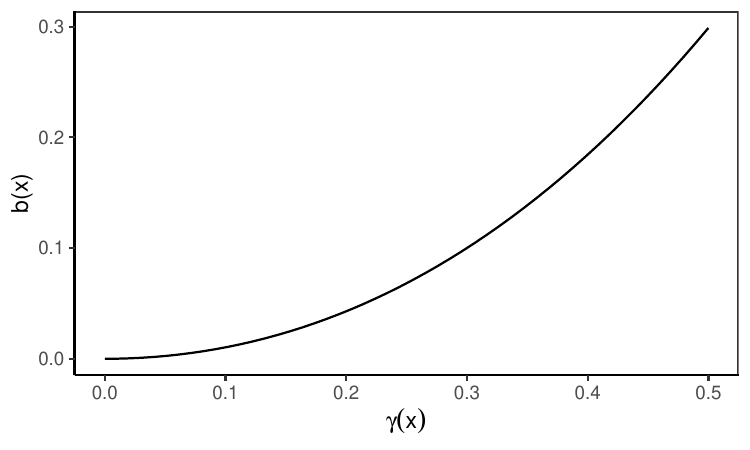}\includegraphics[width=6cm]{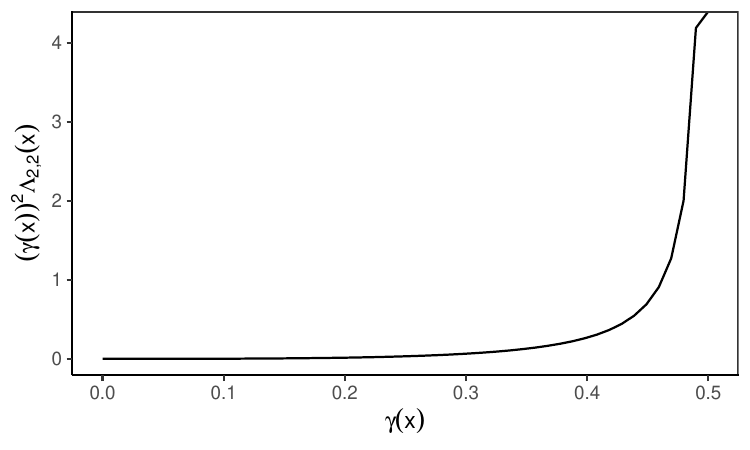}\\
\includegraphics[width=6cm]{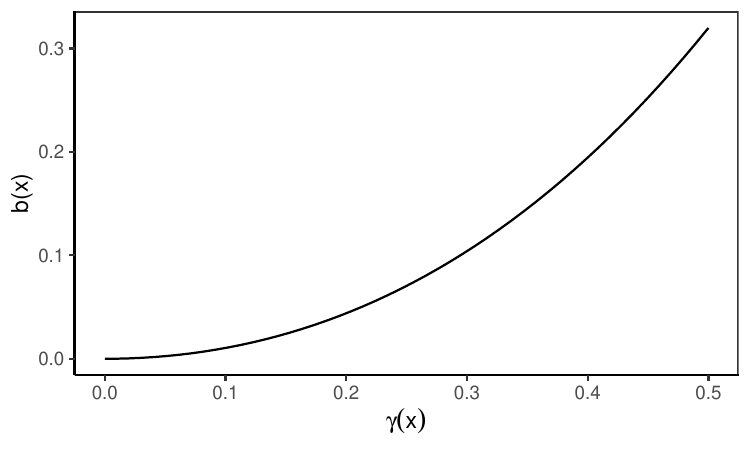}\includegraphics[width=6cm]{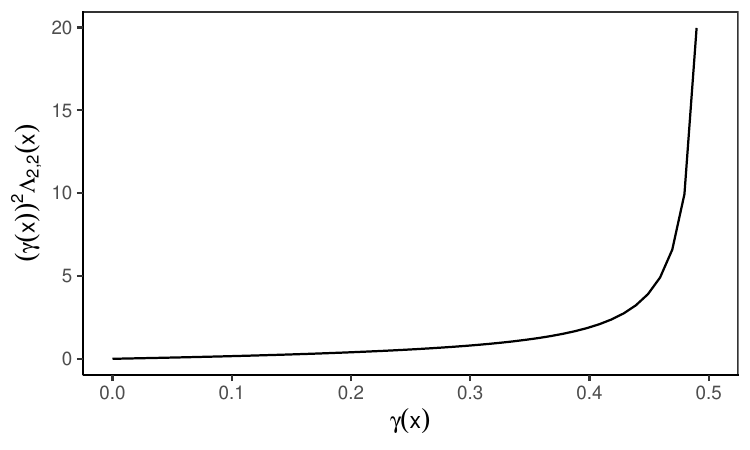}\\
\caption{ Plot of the asymptotic bias term $b(x)$ (left) and
the asymptotic variance term $(\gamma(x))^2\Lambda_{2,2}(x)$ (right) of
the estimator $\widehat{\gamma}_n(\cdot)$
versus the value of $\gamma(x)$
for $J=2$ (top), $J=3$ (bottom).}
\label{Fig3.2}
\end{figure}

The aim of this study is to estimate the ${\RECTM}_{k}(\beta_n|x)$ with
$\beta_n=1-1/n$. From \eqref{eq2.17} and \eqref{eq2.43},
the semi-parametric estimator of ${\RECTM}_{k}(\beta_n|x)$
depends on estimators of the conditional tail index and the
conditional expectile.
Note that computing these estimators requires specification of
the kernel function $K$, the bandwidth $h_n$, the parameter $\alpha_n$ and
a positive decreasing sequence of weights $(\tau_j)_{j\geq1}$.
First, we choose the harmonic sequence $\tau_j=1/j$ for
each $j=1,\ldots,J$ with $J$ a positive integer.
In Figure \ref{Fig3.1}, we show the asymptotic bias term $b(x)$
and the asymptotic variance term $\Lambda_{2,2}(x)$
of $\widehat{\gamma}_n(\cdot)$ with different values of $J$.
It can be seen in this figure that both $b(x)$ and
$\Lambda_{2,2}(x)$ are increasing with respect to $J$.
Moreover, Figure \ref{Fig3.2} illustrates the behavior of the asymptotic variance of $\widehat{\gamma}_n(\cdot)$ as a function of $\gamma(x)$.
As noted in Givens and Hoeting (2013), the shape of the kernel function $K$
is typically not critical, so that the influence of the choice of $K$
on the results is not discussed here and the bi-quadratic kernel function
$K(x)=\frac{15}{16}\left(1-x^2\right)^2{\I}_{\{|x|\leq1\}}$
is considered.
By contrast, the bandwidth has much more influence on the results
than does the kernel function. Now, for the choice of the bandwidth $h_n$,
we use the cross-validation criterion, introduced by Yao (1999), which
generally provides satisfactory results and has been implemented
for instance in Daouia et al. (2011, 2013) and in
Goegebeur et al. (2021a, 2021b, 2021c).
Let $\mathcal{H}$ be a regular grid of size $20$ on the interval
$[h_{\min},h_{\max}]\times \mathcal{R}_X$ with
$h_{\min}=0.05$, $h_{\max}=0.5$ and $\mathcal{R}_X$
the range of $X$, then the parameter $h_n$ is determined
by using the following cross-validation criterion:
\begin{equation*}
  h_{cv}:=\arg\min_{h\in \mathcal{H}}\sum_{i=1}^n\sum_{j=1}^n
  \Big({\I}_{\{Y_i\geq Y_j\}}-\hat{\bar{F}}_{n,-i}(Y_j|X_i)\Big)^2,
\end{equation*}
where $\hat{\bar{F}}_{n,-i}(y|x):=\frac{\sum_{j=1,j\neq i}^n K_h(x-X_j){\I}_{\{Y_j>y\}}}
{\sum_{j=1,j\neq i}^n  K_h(x-X_j)}$.
Once $h_n$ has been selected as $h_{cv}$ by the procedure above,
it thus remains to determine
the value of $\alpha_n$.
Let $x_1, x_2, \cdots,x_{r}$ be regularly distributed on $[0,1]$
and $\mathcal{A}$ be a regular grid of size $20$ on the interval
$[\alpha_{\min},\alpha_{\max}]$ with
$\alpha_{\min}=0.9$ and $\alpha_{\max}=0.96$.
Finally, the parameter $\alpha_n$ is selected by using the
following cross-validation procedure:
\begin{equation*}
  \alpha_n^{*}:=\arg\min_{\alpha_n\in \mathcal{A}}\sum_{t=1}^{r}
  \left(\widetilde{\gamma}_n^{(2)}(x_t)-\widetilde{\gamma}_n^{(3)}(x_t)\right)^2,
\end{equation*}
where $\widetilde{\gamma}_n^{(2)}(\cdot)$ and $\widetilde{\gamma}_n^{(3)}(\cdot)$
are given by \eqref{eq2.51} with $J=2$ and $J=3$, respectively.

\begin{figure}[htbp]
 \centering
\includegraphics[width=6cm]{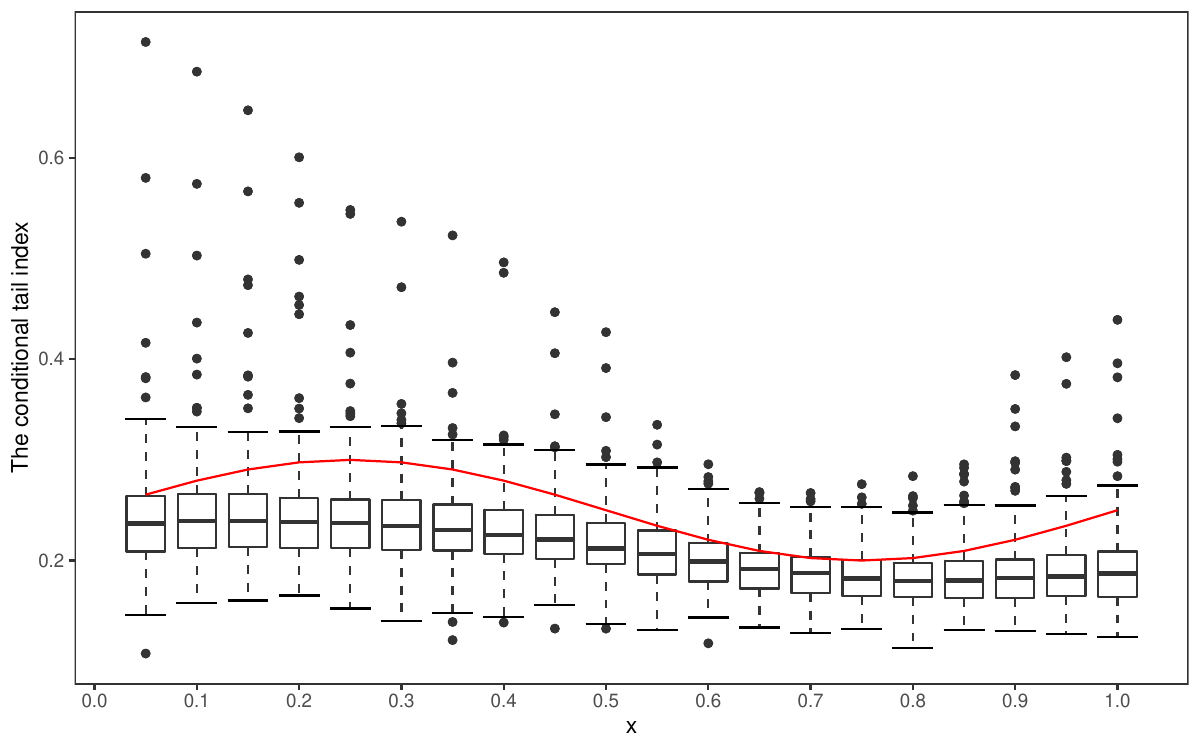}\includegraphics[width=6cm]{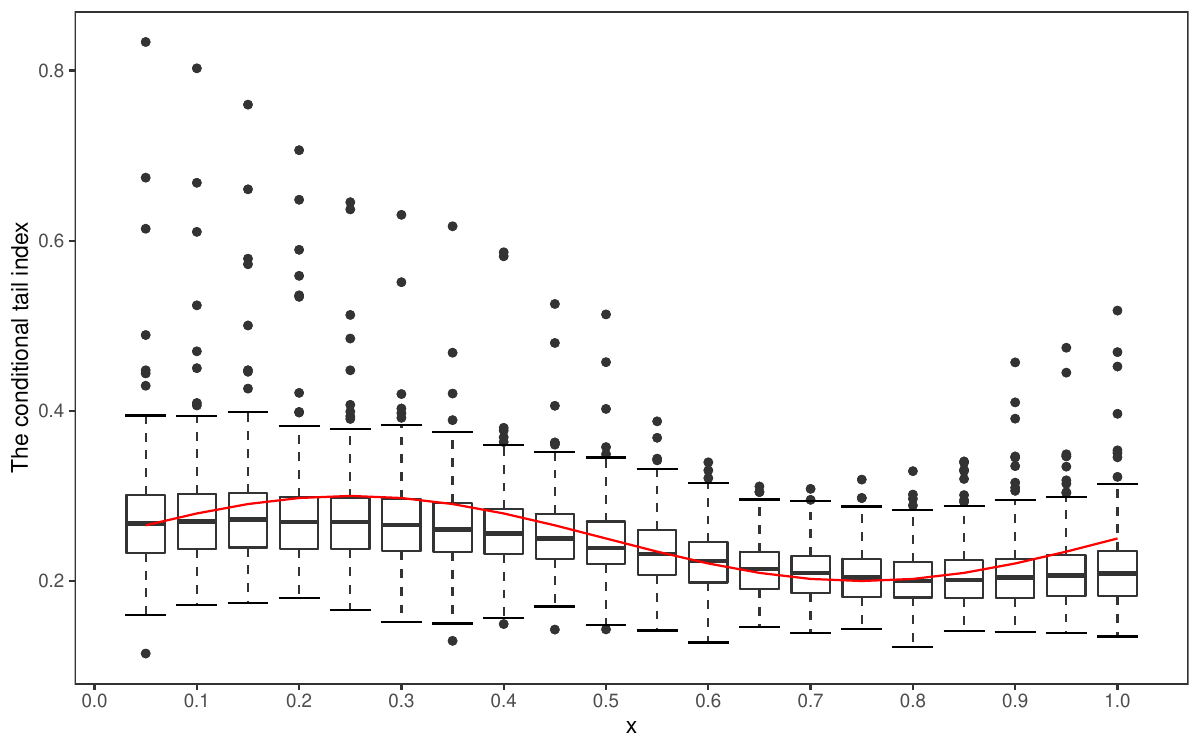}\\
\includegraphics[width=6cm]{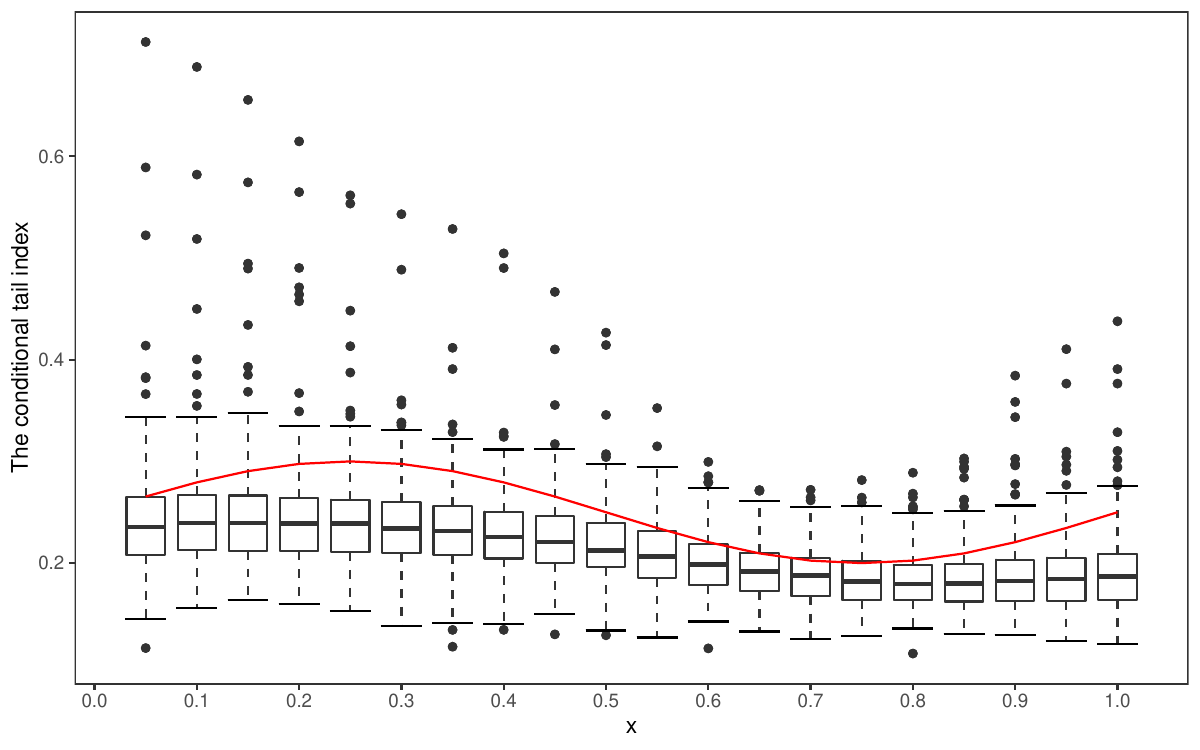}\includegraphics[width=6cm]{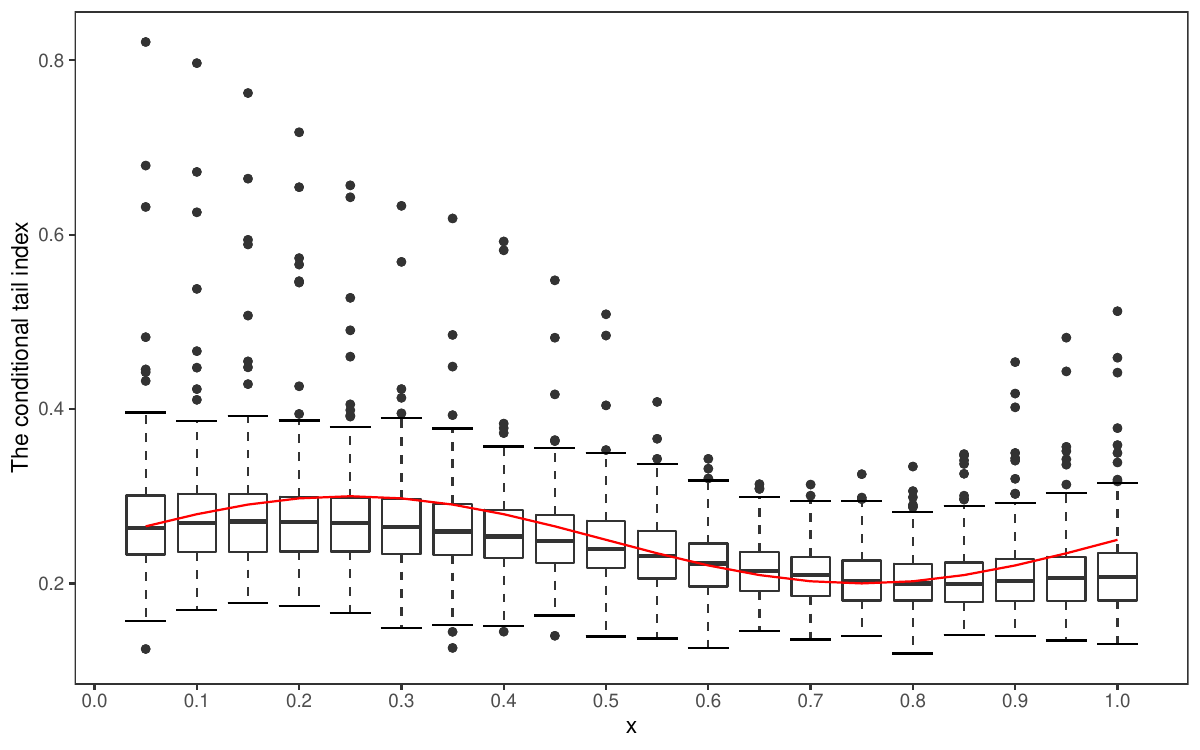}\\
\caption{ Comparison of the true conditional tail index (red curve)
with $\widehat{\gamma}_n(x)$ (left) and $\widetilde{\gamma}_n(x)$ (right)
for $J=2$ (top), $J=3$ (bottom).}
\label{Fig3.4}
\end{figure}

\begin{figure}[htbp]
 \centering
 \includegraphics[width=6cm]{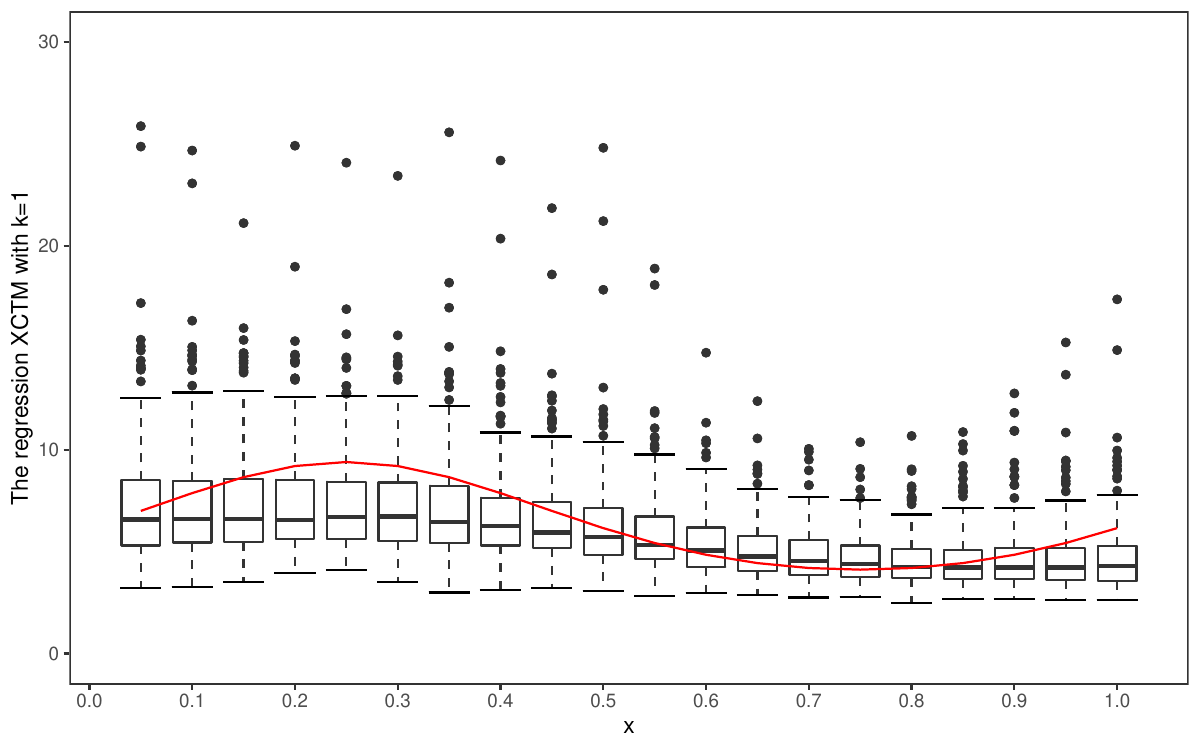}\includegraphics[width=6cm]{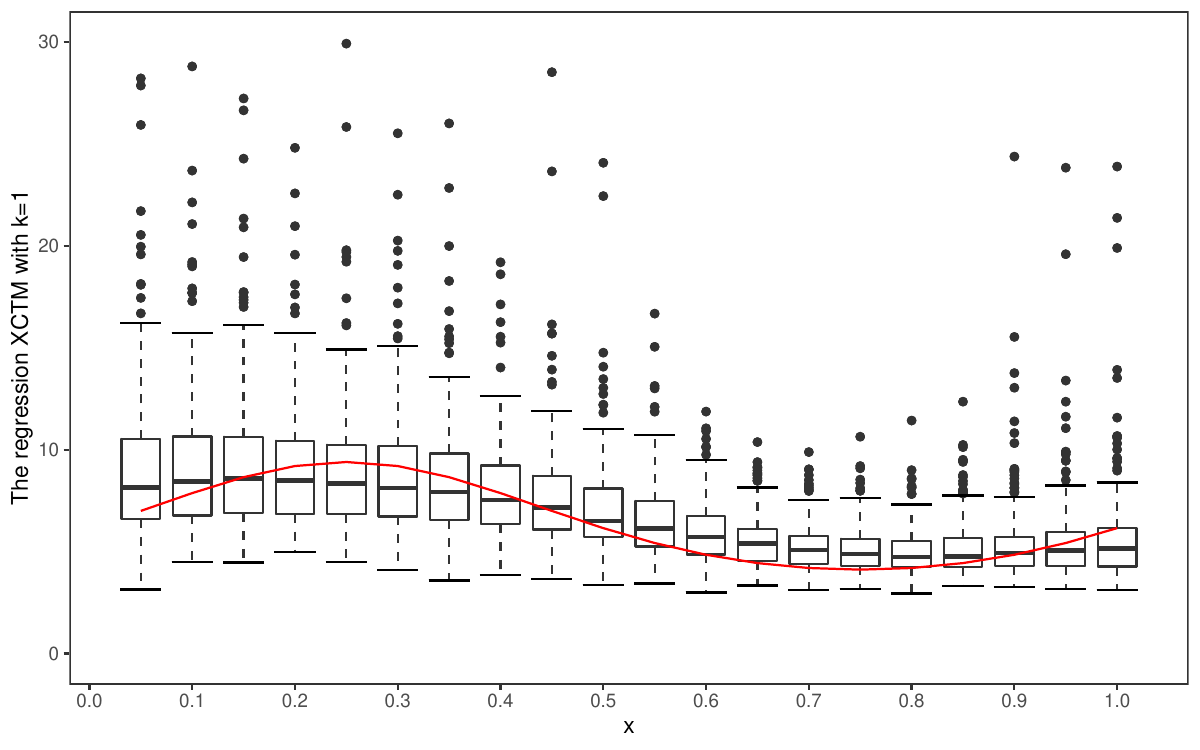}\\
\includegraphics[width=6cm]{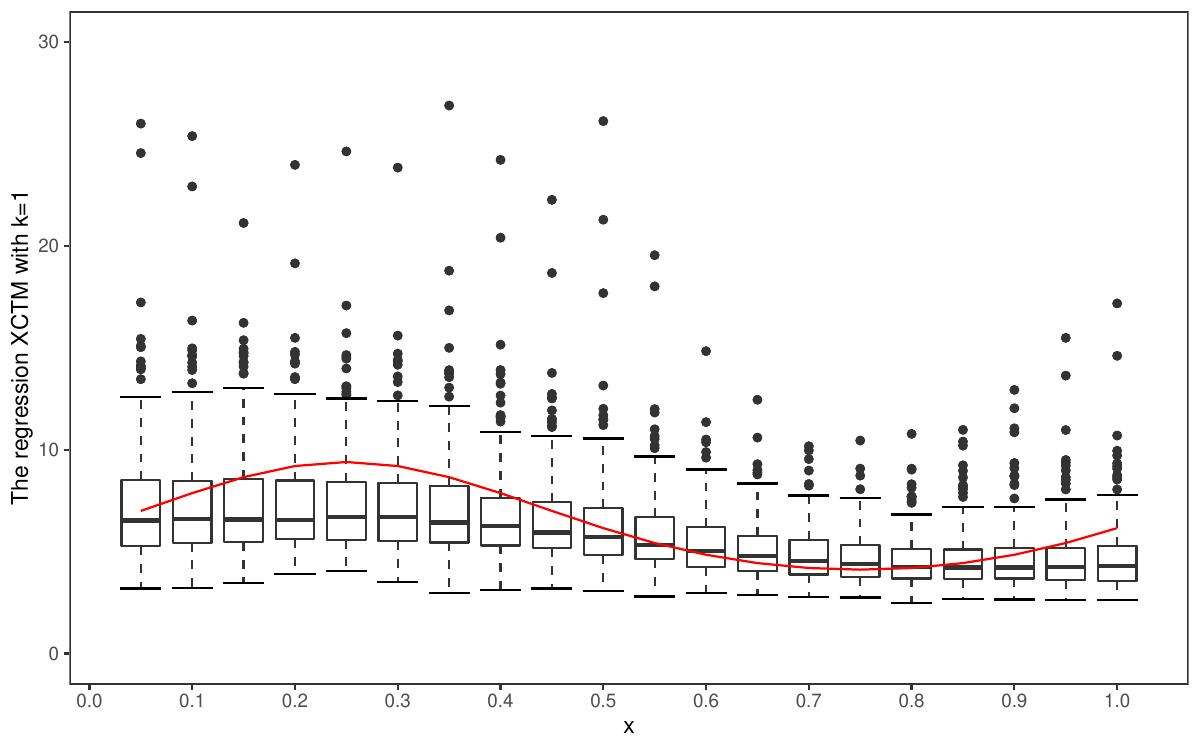}\includegraphics[width=6cm]{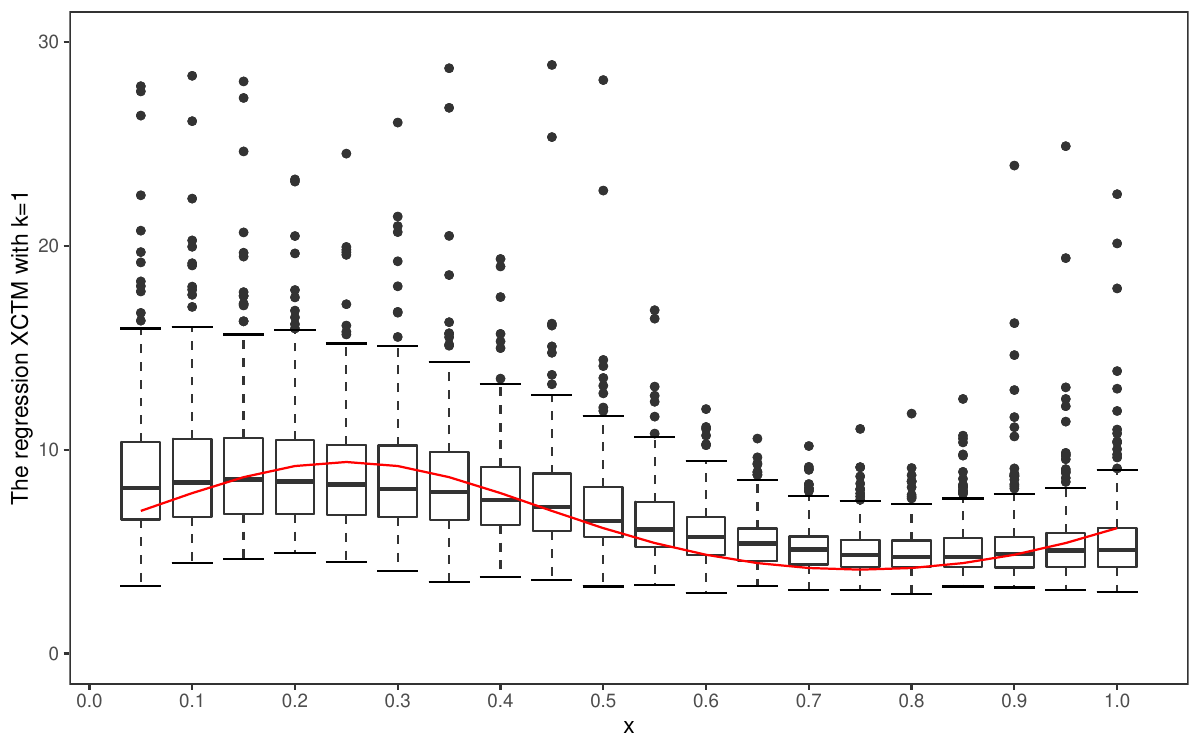}\\
\caption{Comparison of the true
$\RECTM_k(\beta_n|x)$ (red curve) with the nonextrapolated estimator
$\widetilde{\RECTM}_{k,n}(\beta_n|x)$ (left)
and the extrapolated estimator $\widetilde{\RECTM}_{k,n}^W(\beta_n|x)$ (right)
for $J=2$ (top), $J=3$ (bottom).}
\label{Fig3.6}
\end{figure}

We generate $N=200$ replications of an independent sample of size $n=2000$
from a random pair $(X,Y)$. For each sample, we compute our estimators
at $x=0.05,0.1,0.15\ldots,1$.
Figure \ref{Fig3.4} displays the performances of
the simpler estimator $\widehat{\gamma}_n(x)$ and
the bias-reduced estimator $\widetilde{\gamma}_n(x)$
along with the true values of $\gamma(x)$ on boxplots.
In Figure \ref{Fig3.6}, we show the boxplots of
the non-extrapolated estimator $\widetilde{\RECTM}_{k,n}(\beta_n|x)$
and the extrapolated estimator $\widetilde{\RECTM}_{k,n}^W(\beta_n|x)$
for $k=1$, respectively.

In general, the boxplots of our estimators follow the shape of the true values,
although its efficiency depends obviously on the covariate position.
It can be seen in Figure \ref{Fig3.4} that
the bias-reduced estimator performs better than the simpler estimator overall,
while we still get an underestimation of $\gamma(x)$ near the local maxima.
Similar results can also be observed in Figure \ref{Fig3.6},
especially for the nonextrapolated estimator $\widetilde{\RECTM}_{k,n}(\beta_n|x)$,
which can be explained by its high sensitivity to the estimation of $\gamma(x)$.
In other words, making an error in the estimation of $\gamma(x)$ is
more detrimental to the stability of the nonextrapolated estimator.
By contrast, the extrapolated estimator seems to have better performance,
particularly for large values of $\gamma(x)$.
Let us also remark that the represented boxplots tend to
have larger lengths for larger values of $\gamma(x)$.
This result comes from the fact that the variability of each estimator
increases as $\gamma(x)$ increases, as expected in view of the monotonicity
of the asymptotic variances in Theorem \ref{th2.1}, Corollary \ref{cor2.1},
Corollary \ref{cor2.2} and Theorem \ref{th2.3},
also see Figure \ref{Fig3.2}.

\section{Application to health insurance claims data}\label{sec4}
In this section, we consider an application of our methodology
to the {\it health\_insurance} dataset on health insurance claims,
provided in the {\it ExamPAData} R package.
The data set consists of prior year's health insurance claims,
along with patient demographic information, and contains 1338
observations on 7 variables.
Our aim is to estimate extreme expectile-based conditional
tail expectation for the annual medical claims when
the body mass index (BMI) is available.
In this situation, the variable of interest $Y$ is the
annual medical claims, and the covariate $X$ is the BMI.
Duo to the large range of the BMI,
we will focus on the covariate $\log(\text{BMI}) \in [2.9,3.9]$,
which results $n=1332$ pairs $(X_i, Y_i)$ eventually.

\begin{figure}[htbp]
  \centering
 \includegraphics[width=6cm]{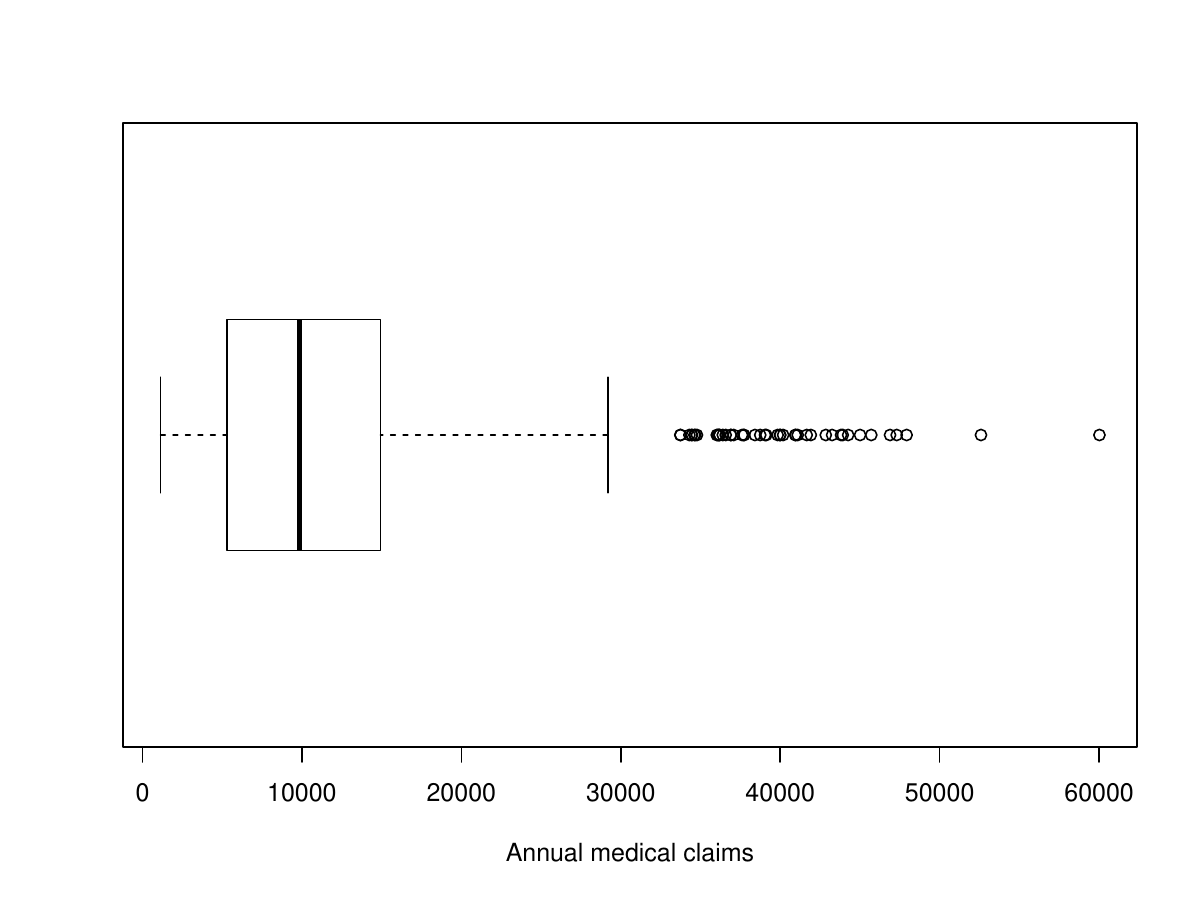}\includegraphics[width=6cm]{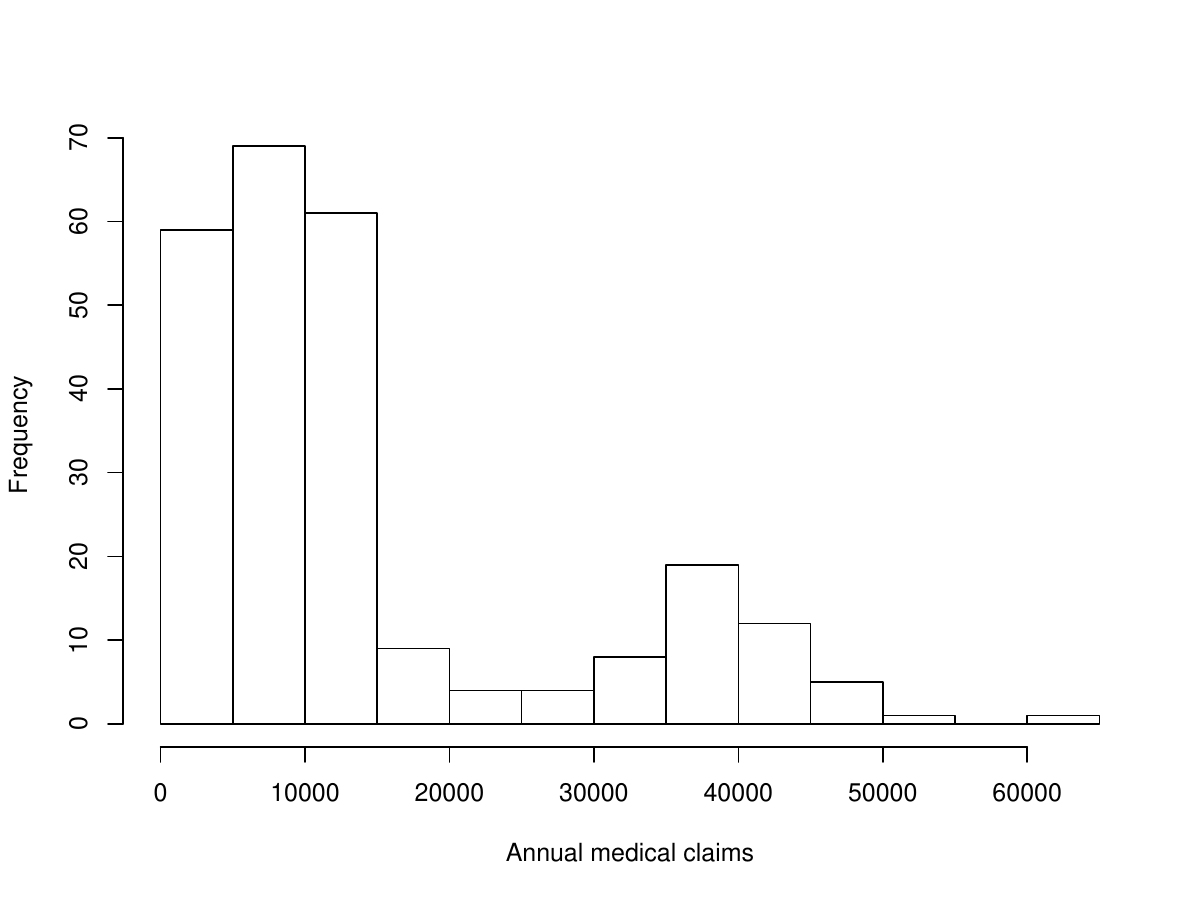}\\
 \includegraphics[width=6cm]{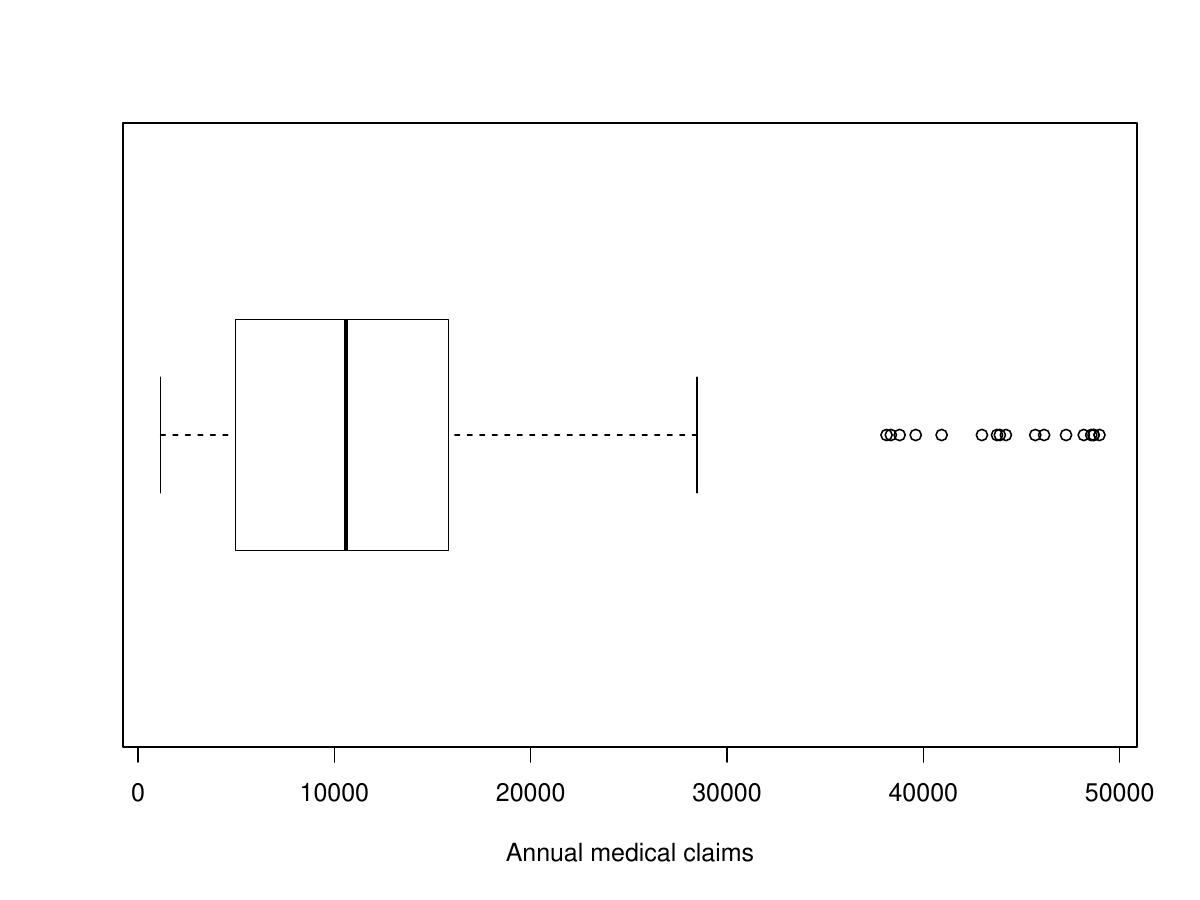}\includegraphics[width=6cm]{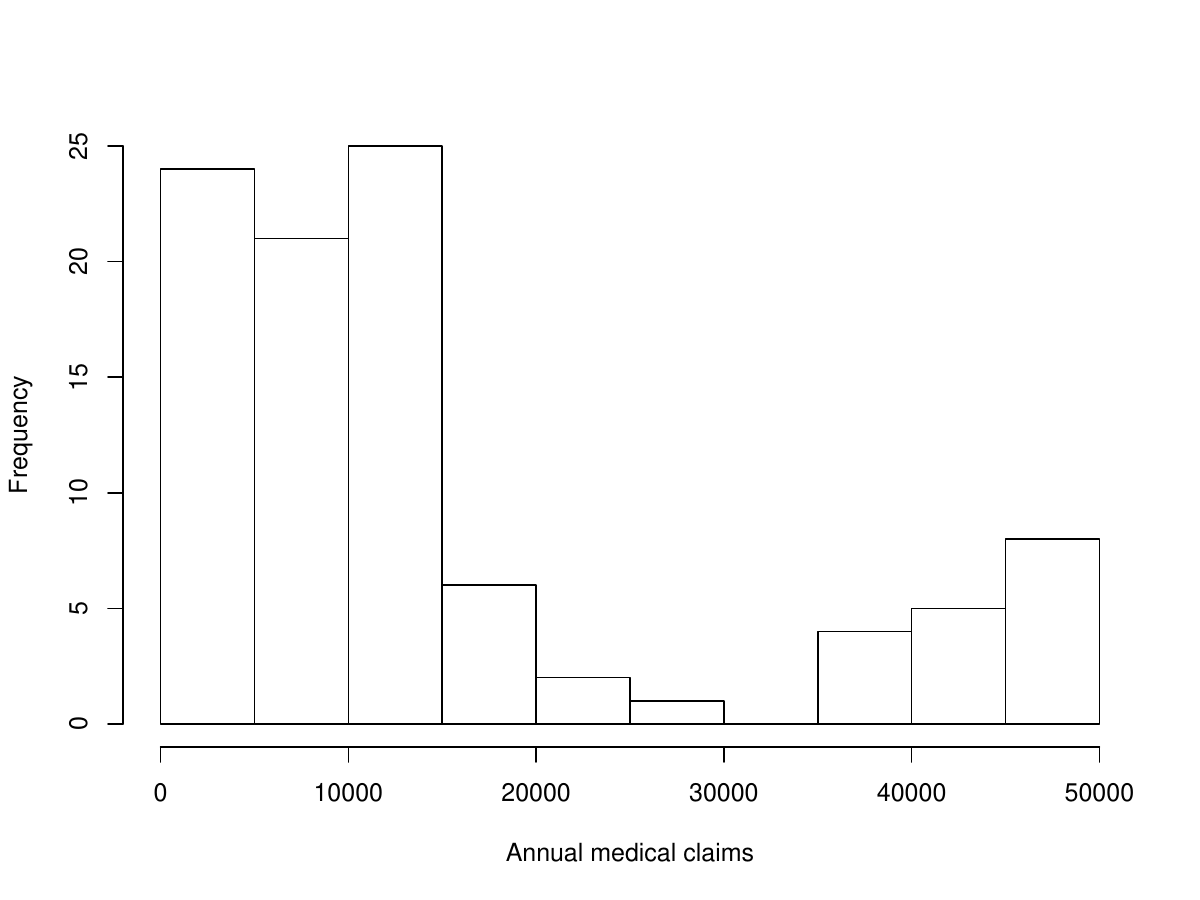}\\
\caption{ Health insurance claims data. Boxplot (left)
and histogram (right) of annual medical claims at
$\log(\text{BMI})=3.5$ (top) and $\log(\text{BMI})=3.7$ (bottom).}
\label{Fig4.2}
\end{figure}

\begin{figure}[htbp]
  \centering
\includegraphics[width=6cm]{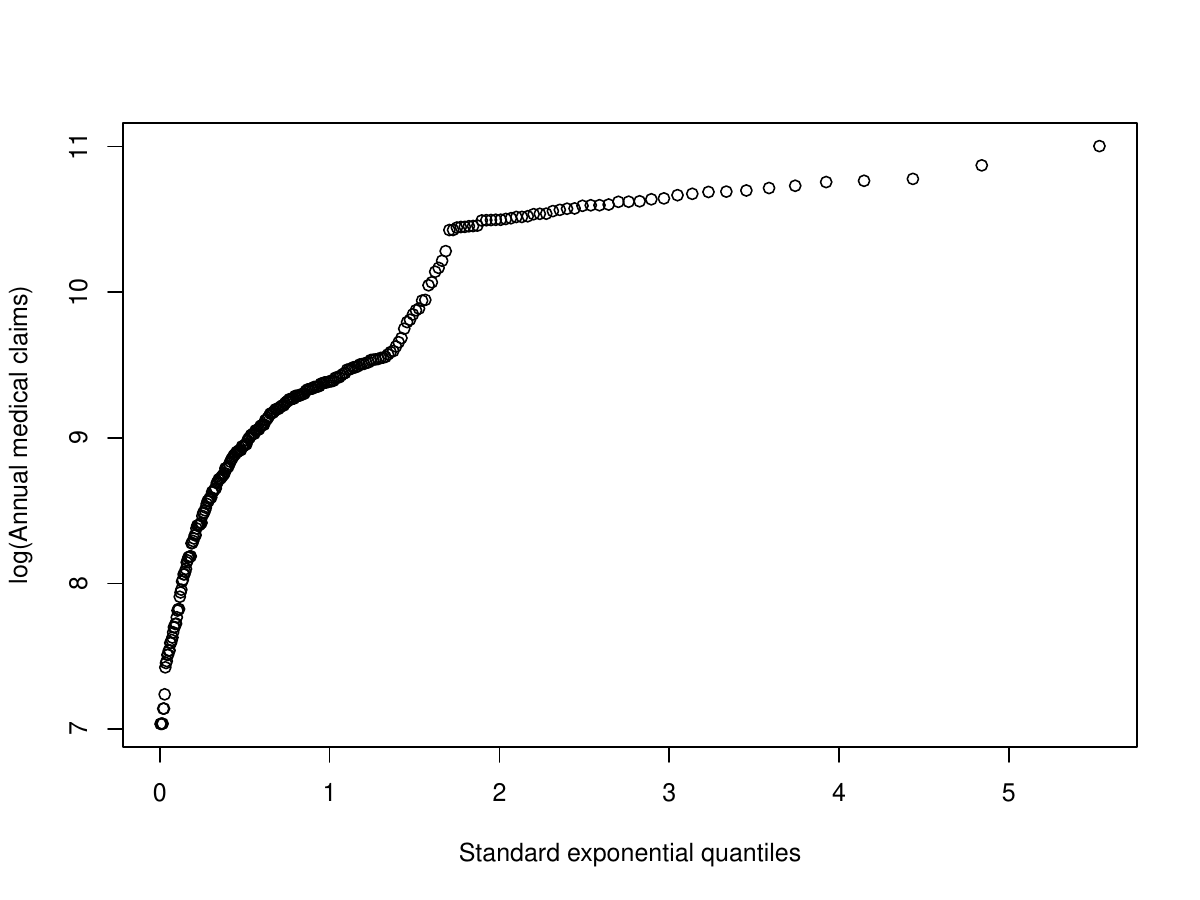}\includegraphics[width=6cm]{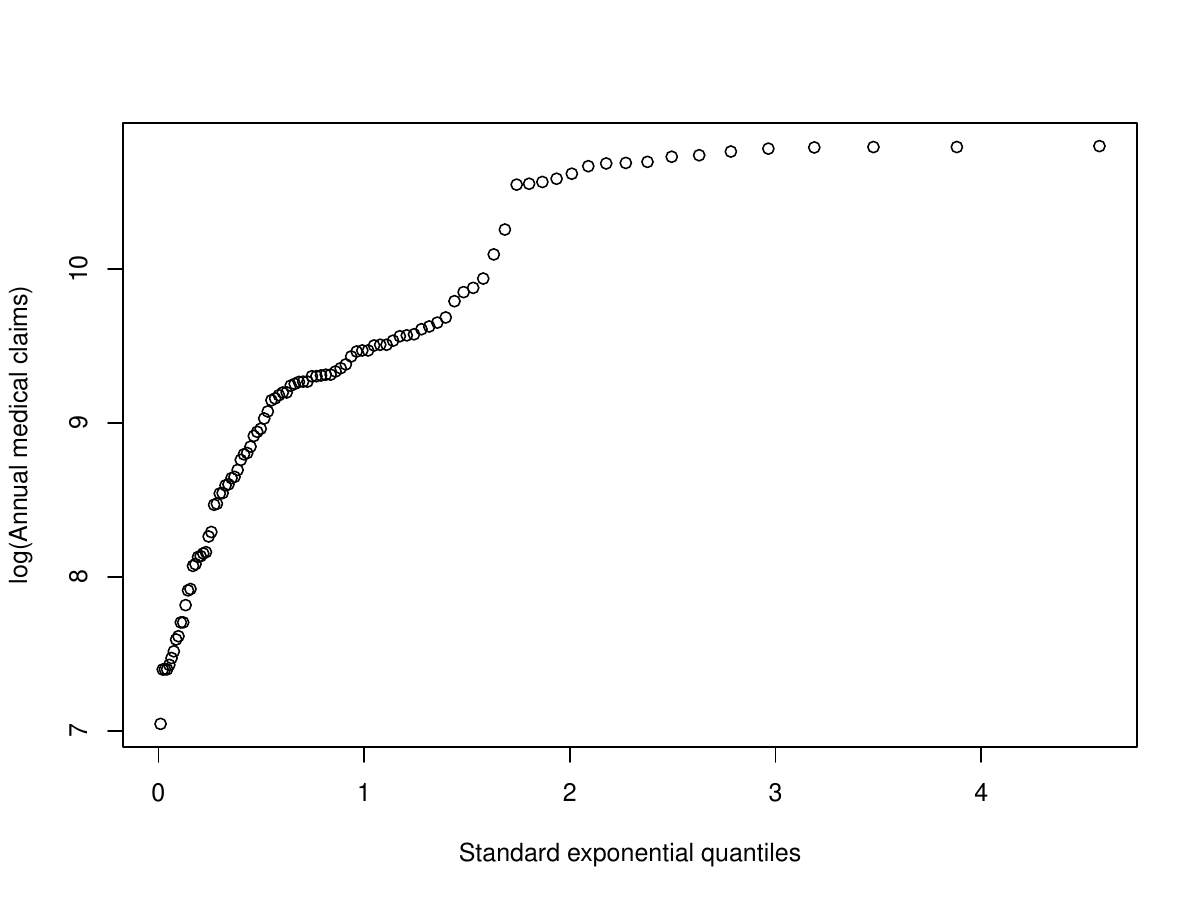}\\
\caption{Health insurance claims data. Local Pareto QQ-plot
of annual medical claims at $\log(\text{BMI})=3.5$ (left)
and $\log(\text{BMI})=3.7$ (right).}
\label{Fig4.3}
\end{figure}

In Figure \ref{Fig4.2}, we provide the boxplots and histograms of
annual medical claims for which the covariate is
in a neighborhood of $X=3.5$ and $X=3.7$.
It can be seen in this figure that the annual medical claims
seem to have a heavy right tail.
To further validate the assumption of underlying conditional
heavy-tailed distributions for the annual medical claims,
we show local Pareto quantile plots, see Figure \ref{Fig4.3}.
Clearly, the local Pareto quantile plots become
approximately linear near the largest observations,
which indicates that the heavy-tailed framework makes sense.

\begin{figure}[htbp]
  \centering
 \includegraphics[width=6cm]{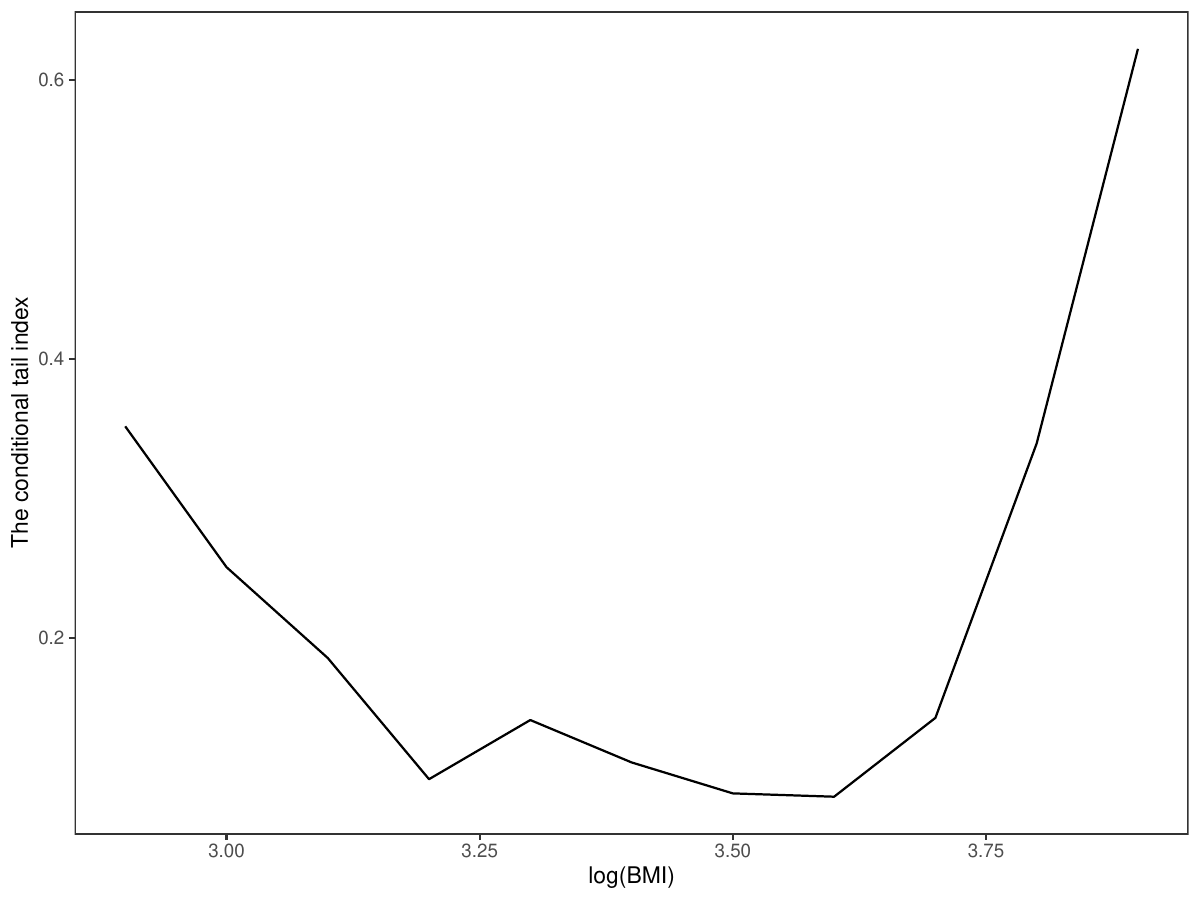}\includegraphics[width=6cm]{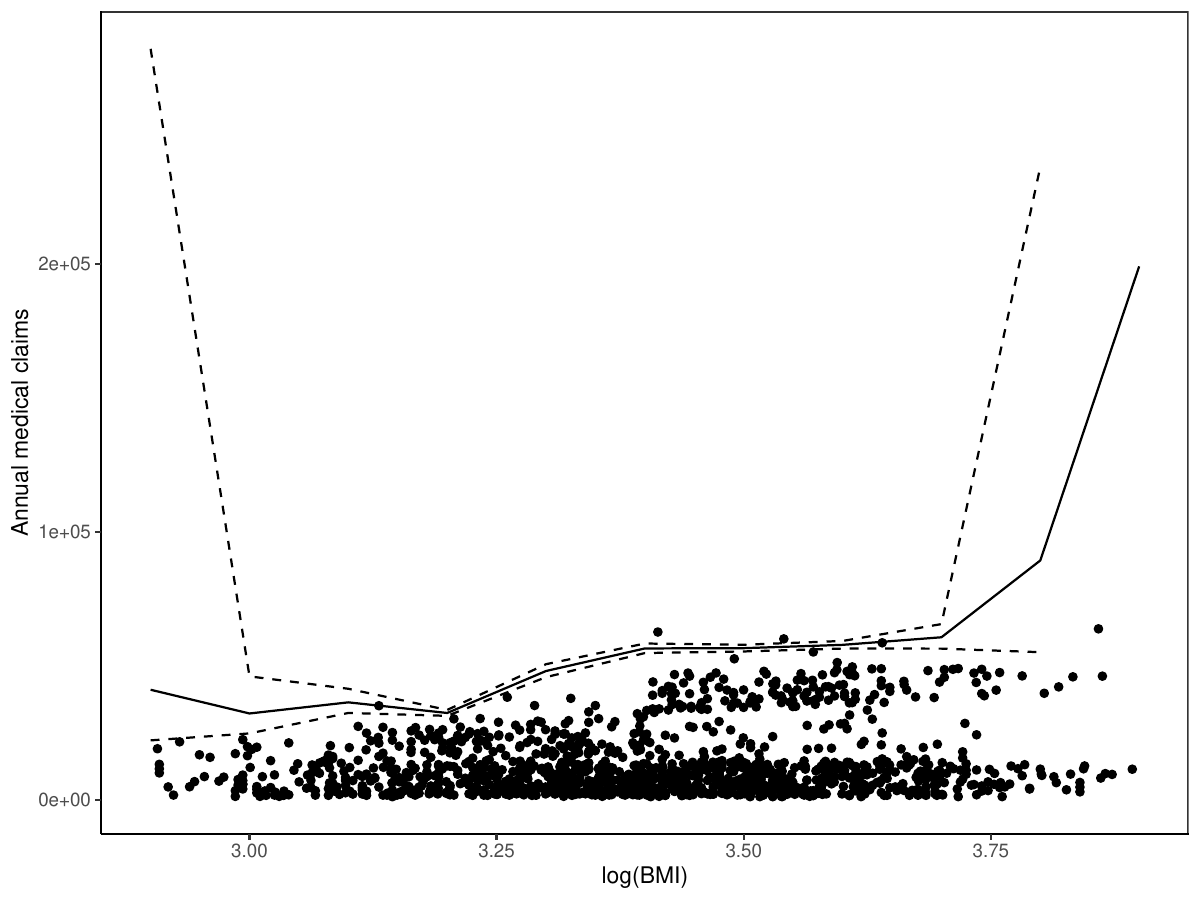}\\
\caption{Health insurance claims data.
Left: $\widetilde\gamma(x)$ versus $x$.
Right: $\widetilde{\RECTM}_{k,n}^W(\beta_n|x)$ with
$k=1$ and their pointwise $95\%$ confidence intervals (dashed line).}
\label{Fig4.4}
\end{figure}

Here, we consider the bi-quadratic kernel function
$K(x)=\frac{15}{16}\left(1-x^2\right)^2\I_{\{|x|\leq1\}}$.
Using this kernel function, the cross-validation procedure
introduced in Section \ref{sec3} yields $h_{cv}\approx 0.12$
and $\alpha_n^{*}\approx 0.96$.
Choosing the harmonic sequence $\tau_j=1/j$ for each $j=1,2$,
Figure \ref{Fig4.4} (left) displays the estimates
$\widetilde{\gamma}_n(x)$ as a function of $x$.
Finally, we focus on the estimation of ${\RECTM}_k(\beta_n|x)$ with
$k=1$ and $\beta_n=1-5/n\approx0.996$, and the estimates
are represented in Figure \ref{Fig4.4} (right).
Using the asymptotic normality of $\widetilde{\RECTM}_{k,n}^W(\beta_n|x)$
for $k=1$ in Theorem \ref{th2.3}, pointwise $95\%$
Gaussian asymptotic confidence intervals can also be computed.
Let $z_{1-\theta/2}$ denote the $(1-\theta/2)$-quantile of the standard normal
distribution $\N(0,1)$, then the $(1-\theta)$-confidence intervals are given by
\begin{equation*}
  \left(
  \frac{\widetilde{\RECTM}_{k,n}^W(\beta_n|x)}
  {1+z_{1-\theta/2}\frac{k||K||_2\widetilde{\gamma}_n(x)\log((1-\alpha_n)/(1-\beta_n))
  \sqrt{\hat{\Lambda}_{2,2}(x)}}
  {\sqrt{nh_n^p(1-\alpha_n)\hat{g}_n(x)}}},
  \frac{\widetilde{\RECTM}_{k,n}^W(\beta_n|x)}
  {1-z_{1-\theta/2}\frac{k||K||_2\widetilde{\gamma}_n(x)\log((1-\alpha_n)/(1-\beta_n))
  \sqrt{\hat{\Lambda}_{2,2}(x)}}
  {\sqrt{nh_n^p(1-\alpha_n)\hat{g}_n(x)}}}
  \right),
\end{equation*}
where $\hat{\Lambda}_{2,2}(x)$ is an estimator of $\Lambda_{2,2}(x)$ obtained by
plugging in the estimator $\widetilde{\gamma}_n(x)$ in place of $\gamma(x)$,
$\widetilde{\gamma}_n(x)$ and $\hat{g}_n(x)$ are
estimates for $\gamma(x)$ and $g(x)$, respectively.

One can see that we obtain larger estimates for
the conditional tail index $\gamma(x)$
when the values of $\log(\text{BMI})$ is too large or too small.
This result seems reasonable in practice since one may be
in the state of sub-health when the values of $\log(\text{BMI})$
is too large or too small, which increases the probability of large claims.
Obviously, the confidence intervals are wider in some areas where
the estimates for $\gamma(x)$ is large.
This can be expected because of the fact that the variability
of the estimator increases as $\gamma(x)$ increases.
At the last covariate position, we could not obtain a confidence interval
due to a negative estimate of $\Lambda_{2,2}(x)$.

\section{Proof}\label{sec5}
In order to prove our main results, some auxiliary lemmas are needed.
The first one provides the second order asymptotic expansion of $e(\cdot|x)$,
which is similar to that in Daouia et al. (2020) but under slightly different assumptions.

\begin{lemma}\label{lem5.1}
Assume that  the second-order condition $(C2)$ holds.
Suppose also that $\gamma(x)<1$ and $\E[Y_{-}|X=x]<\infty$.
Let $\alpha_n\rightarrow1$, $\beta_n\rightarrow1$ be such that
$\frac{1-\beta_n}{1-\alpha_n}\rightarrow\tau\in\R$ as $n\rightarrow\infty$.
Then
\begin{equation*}
  \left(\frac{1-\beta_n}{1-\alpha_n}\right)^{\gamma(x)}\frac{e(\beta_n|x)}{e(\alpha_n|x)}
  =1+\frac{\gamma(x)}{e(\alpha_n|x)}
  (\tau^{\gamma(x)}-1)
  \big(m^{(1)}(x)+o(1)\big)
  +O\big(A((1-\alpha_n)^{-1}|x)\big).
\end{equation*}
\end{lemma}

\noindent
{\bf Proof. }
Write
\begin{equation*}
  \frac{e(\beta_n|x)}{e(\alpha_n|x)}
  =\frac{e(\beta_n|x)}{q(\beta_n|x)}
  \times\frac{q(\alpha_n|x)}{e(\alpha_n|x)}
  \times\frac{q(\beta_n|x)}{q(\alpha_n|x)}.
\end{equation*}
According to Proposition 1 in Daouia et al. (2020), we get
\begin{equation*}
  \frac{e(\beta_n|x)}{q(\beta_n|x)}
  =(1/\gamma(x)-1)^{-\gamma(x)}
  \left(1+\frac{\gamma(x)}{e(\beta_n|x)}\big(m^{(1)}(x)+o(1)\big)
  +O\big(A((1-\beta_n)^{-1}|x)\big)\right)
\end{equation*}
and
\begin{equation*}
  \frac{q(\alpha_n|x)}{e(\alpha_n|x)}
  =(1/\gamma(x)-1)^{\gamma(x)}
  \left(1-\frac{\gamma(x)}{e(\alpha_n|x)}\big(m^{(1)}(x)+o(1)\big)
  +O\big(A((1-\alpha_n)^{-1}|x)\big)\right).
\end{equation*}
Moreover, the second-order regularly varying condition implies that
\begin{equation*}
  \frac{q(\beta_n|x)}{q(\alpha_n|x)}
  =\left(\frac{1-\beta_n}{1-\alpha_n}\right)^{-\gamma(x)}
  \left(1+O\big(A((1-\alpha_n)^{-1}|x)\big)\right).
\end{equation*}
Recall that $e(1-1/\cdot|x)$ and $A(\cdot|x)$ are regularly varying at infinity
with respective indices $\gamma(x)>0$ and $\rho(x)\leq0$, we have
\begin{equation*}
  \frac{e(\beta_n|x)}{e(\alpha_n|x)}
  =\left(\frac{1-\beta_n}{1-\alpha_n}\right)^{-\gamma(x)}
  \left(1+\frac{\gamma(x)}{e(\alpha_n|x)}
  (\tau^{\gamma(x)}-1)
  \big(m^{(1)}(x)+o(1)\big)
  +O\big(A((1-\alpha_n)^{-1}|x)\big)\right),
\end{equation*}
and hence the desired result follows immediately.
\qed

The second lemma comes from Lemma 5.4 in Xiong and Peng (2023).
We refer to Pan et al. (2013, Theorem 4.2) for a similar result
in the unconditional case.

\begin{lemma}\label{lem5.4}
Assume that  the second-order condition $(C2)$ holds.
Then, for $k\in(0,1/\gamma(x))$ we have
\begin{equation*}
  \lim_{t\rightarrow\infty}{\frac{1}{A(1/\bar{F}(t|x)|x)}}\left\{\frac{\E[Y^k|Y>t,X=x]}{t^k}-
  \frac{1/\gamma(x)}{1/\gamma(x)-k}\right\}
  =\frac{k}{(1/\gamma(x)-k)\left(1/\gamma(x)-k-\rho(x)\right)}.
\end{equation*}
\end{lemma}

Using the results of Lemma \ref{lem5.1} and Lemma \ref{lem5.4},
the second order asymptotic expansion of ${\RECTM}_k(\cdot|x)$
is established.

\begin{lemma}\label{lem5.2}
Under the assumptions of Lemma \ref{lem5.1}, we have for $k\in(0,1/\gamma(x))$,
{\footnotesize\begin{equation*}
  \left(\frac{1-\beta_n}{1-\alpha_n}\right)^{k\gamma(x)}
  \frac{{\RECTM}_k(\beta_n|x)}{{\RECTM}_k(\alpha_n|x)}
  =1+\frac{k\gamma(x)}{e(\alpha_n|x)}
  (\tau^{\gamma(x)}-1)
  \big(m^{(1)}(x)+o(1)\big)
  +O\big(A((1-\alpha_n)^{-1}|x)\big)
  +{O\big(A(1/\bar{F}(e(\alpha_n|x)|x)|x)\big)}.
\end{equation*}}
\end{lemma}

\noindent
{\bf Proof. }
Note that
\begin{equation*}
  \frac{{\RECTM}_k(\beta_n|x)}{{\RECTM}_k(\alpha_n|x)}
  =\frac{{\RECTM}_k(\beta_n|x)}{(e(\beta_n|x))^k}
  \times\frac{(e(\alpha_n|x))^k}{{\RECTM}_k(\alpha_n|x)}
  \times\left(\frac{e(\beta_n|x)}{e(\alpha_n|x)}\right)^k.
\end{equation*}
Applying Lemma \ref{lem5.4} with $t=e(\beta_n|x)$ and $t=e(\alpha_n|x)$ we have
\begin{equation*}
  \frac{{\RECTM}_k(\beta_n|x)}{(e(\beta_n|x))^k}
  =(1-k\gamma(x))^{-1}\left(1+{O\big(A(1/\bar{F}(e(\beta_n|x)|x)|x)\big)}\right)
\end{equation*}
and
\begin{equation*}
  \frac{(e(\alpha_n|x))^k}{{\RECTM}_k(\alpha_n|x)}
  =(1-k\gamma(x))\left(1+{O\big(A(1/\bar{F}(e(\alpha_n|x)|x)|x)\big)}\right).
\end{equation*}
Moreover, it follows from Lemma \ref{lem5.1} that
\begin{equation*}
  \left(\frac{e(\beta_n|x)}{e(\alpha_n|x)}\right)^k
  =\left(\frac{1-\beta_n}{1-\alpha_n}\right)^{-k\gamma(x)}
  \left(1+\frac{k\gamma(x)}{e(\alpha_n|x)}
  (\tau^{\gamma(x)}-1)
  \big(m^{(1)}(x)+o(1)\big)
  +O\big(A((1-\alpha_n)^{-1}|x)\big)\right).
\end{equation*}
Thus, by noticing that $e(1-1/\cdot|x)$ and $A(\cdot|x)$ are regularly varying at
infinity with respective indices $\gamma(x)>0$ and $\rho(x)\leq0$, we have
{\scriptsize\begin{equation*}
  \frac{{\RECTM}_k(\beta_n|x)}{{\RECTM}_k(\alpha_n|x)}
  =\left(\frac{1\!-\!\beta_n}{1\!-\!\alpha_n}\right)^{\!-k\gamma(x)}
  \left(1\!+\!\frac{k\gamma(x)}{e(\alpha_n|x)}
  (\tau^{\gamma(x)}\!\!\!-\!1)
  \big(m^{(1)}(x)\!+\!o(1)\big)
  \!+\!O\big(A((1\!-\!\alpha_n)^{-1}|x)\big)
  \!+\!{O\big(A(1/\bar{F}(e(\alpha_n|x)|x)|x)\big)}\right).
\end{equation*}}
Hence, the desired result follows immediately.
\qed

The last lemma provides the asymptotic properties of the estimator
$\hat{e}_n(\alpha_n|x)$ due to Theorem 1 in Girard et al. (2022).

\begin{lemma}\label{lem5.3}
Assume that $(C1)$, $(C3)$ and $(C4)$ hold. Suppose also
that $\gamma(x)<1/2$ and that there exists $\delta\in(0,1)$ with
$\E[Y_{-}^{2+\delta}|X=x]<\infty$.
Let $\alpha_n\rightarrow1$, $h_n\rightarrow0$ be such that
$n{h_n}^p(1-\alpha_n)\rightarrow\infty$,
$n{h_n}^{p+2}(1-\alpha_n)\rightarrow0$,
$\sqrt{n{h_n}^p(1-\alpha_n)}\log(1-\alpha_n)
\times\omega_{h_n}((1-\delta)e(\alpha_n|x)|x)\rightarrow0$.
Then
\begin{equation*}
  \sqrt{nh_n^p(1-\alpha_n)}
  \left\{
  \left(\frac{\hat{e}_n(\alpha_{n,j}|x)}{e(\alpha_{n,j}|x)}-1\right)_{1\leq j\leq J}
  \right\}^T
  \overset{d}{\longrightarrow}
  \N\left(0,\frac{\Vert K\Vert_2^2\gamma^2(x)}{g(x)}\tilde{\Lambda}(x)\right),
\end{equation*}
where $1-\alpha_{n,j}=a_j(1-\alpha_n)$ for some $0<a_1<a_2<\cdots<a_J\leq1$ and
$\tilde{\Lambda}(x)$ is the symmetric matrix having entries
$\tilde{\Lambda}_{j,l}(x)=a^{-1}_l\big(\frac{1}{1-2\gamma(x)}(a_j/a_l)^{-\gamma(x)}-1\big)$,
for $j,l\in\{1,\ldots,J\}$, $j\leq l$.
\end{lemma}

\noindent
{\bf Proof of Theorem \ref{th2.1}.}
Write $c_J=\sum_{j=1}^J\log(1/\tau_j)$, then by \eqref{eq2.18} we have
{\footnotesize\begin{equation*}
\hat{\gamma}_n(x)-\gamma(x)=c_J^{-1}\sum_{j=1}^J\left(
\log\left(\frac{\hat{e}_n(1-\tau_j(1-\alpha_n)|x)}{e(1-\tau_j(1-\alpha_n)|x)}\right)
-\log\left(\frac{\hat{e}_n(\alpha_n|x)}{e(\alpha_n|x)}\right)\right)
+c_J^{-1}\sum_{j=1}^J\log\left(
\tau_j^{\gamma(x)}\left(\frac{e(1-\tau_j(1-\alpha_n)|x)}{e(\alpha_n|x)}\right)\right).
\end{equation*}}
Let $\beta=(\beta_1,\beta_2)^T\neq0$ in $\R^2$, then we consider the decomposition
\begin{equation}\label{eq5.1}
  \beta_1\sqrt{nh_n^p(1-\alpha_n)}\left(\frac{\hat{e}_n(\alpha_n|x)}{e(\alpha_n|x)}-1\right)
  +\beta_2\sqrt{nh_n^p(1-\alpha_n)}\left(\hat{\gamma}_n(x)-\gamma(x)\right)
  = c_J^{-1}\left(I_{n,1}+I_{n,1}\right),
\end{equation}
where
{\footnotesize\begin{eqnarray*}
  I_{n,1} &=& \beta_1c_J\sqrt{nh_n^p(1-\alpha_n)}
  \left(\frac{\hat{e}_n(\alpha_n|x)}{e(\alpha_n|x)}-1\right)
  +\beta_2\sqrt{nh_n^p(1-\alpha_n)}\sum_{j=1}^J
  \left(
  \log\left(\frac{\hat{e}_n(1-\tau_j(1-\alpha_n)|x)}{e(1-\tau_j(1-\alpha_n)|x)}\right)
  -\log\left(\frac{\hat{e}_n(\alpha_n|x)}{e(\alpha_n|x)}\right)
  \right),\\
  I_{n,2} &=& \beta_2\sqrt{nh_n^p(1-\alpha_n)}\sum_{j=1}^J \log
  \left(
  \tau_j^{\gamma(x)}\left(\frac{e(1-\tau_j(1-\alpha_n)|x)}{e(\alpha_n|x)}\right)
  \right).
\end{eqnarray*}}
Note that a direct consequence of Lemma \ref{lem5.3} is that
$\frac{\hat{e}_n(1-\tau_j(1-\alpha_n)|x)}
{e(1-\tau_j(1-\alpha_n)|x)}\overset{\P}{\rightarrow}1$
as $n\rightarrow\infty$. This implies that for large $n$,
{\scriptsize\begin{equation*}
  \left\{\sqrt{nh_n^p(1-\alpha_n)}
  \log\left(\frac{\hat{e}_n(1-\tau_j(1-\alpha_n)|x)}{e(1-\tau_j(1-\alpha_n)|x)}\right)
  \right\}_{j=1,\cdots,J}
  =(1+o_{\P}(1))\left\{\sqrt{nh_n^p(1-\alpha_n)}
  \left(\frac{\hat{e}_n(1-\tau_j(1-\alpha_n)|x)}{e(1-\tau_j(1-\alpha_n)|x)}-1\right)
  \right\}_{j=1,\cdots,J}.
\end{equation*}}
Thus, using Lemma \ref{lem5.3} again we have
\begin{equation*}
  c_J^{-1}I_{n,1}\overset{d}{\longrightarrow}
  \N\left(0,\frac{\Vert K\Vert_2^2\gamma^2(x)}{g(x)c_J^2}
  \theta^T\bar{\Lambda}(x)\theta\right),
\end{equation*}
where $\theta=(\beta_1c_J+\beta_2(1-J),\beta_2,\cdots,\beta_2)^T\in\R^J$
and $\bar{\Lambda}(x)$ is the symmetric matrix having entries
$\bar{\Lambda}_{j,l}(x)=\tau^{-1}_j\big(\frac{1}{1-2\gamma(x)}
(\tau_l/\tau_j)^{-\gamma(x)}-1\big)$, for $j,l\in\{1,\ldots,J\}$, $j\leq l$.
An elementary computation gives $\theta^T\bar{\Lambda}(x)\theta=c_J^2\beta^T\Lambda(x)\beta$,
and hence
\begin{equation}\label{eq5.2}
  c_J^{-1}I_{n,1}\overset{d}{\longrightarrow}
  \N\left(0,\frac{\Vert K\Vert_2^2\gamma^2(x)}{g(x)}\beta^T\Lambda(x)\beta\right).
\end{equation}
On the other hand, applying Lemma \ref{lem5.1} and the fact that
$A(t|x)\rightarrow0$ as $t\rightarrow\infty$, we have
{\small\begin{equation}\label{eq5.3}
  c_J^{-1}I_{n,2}= c_J^{-1}\beta_2\gamma(x)
  \frac{\sqrt{nh_n^p(1-\alpha_n)}}{e(\alpha_n|x)}\big(m^{(1)}(x)+o(1)\big)
  \sum_{j=1}^J(\tau_j^{\gamma(x)}-1)
  +O\left(\sqrt{nh_n^p(1-\alpha_n)}A((1-\alpha_n)^{-1}|x)\right).
\end{equation}}
Therefore, the result follows by combining \eqref{eq5.1}-\eqref{eq5.3}
and using the conditions
$\frac{\sqrt{nh_n^p(1-\alpha_n)}}{e(\alpha_n|x)}\rightarrow\lambda(x)$
and
$\sqrt{nh_n^p(1-\alpha_n)}A((1-\alpha_n)^{-1}|x)\rightarrow0$.
\qed

\noindent
{\bf Proof of Theorem \ref{th2.2}.}
From \eqref{eq2.17}, we have
\begin{equation}\label{eq5.4}
  \frac{\widetilde{\RECTM}_{k,n}(\alpha_n|x)}{\RECTM_k(\alpha_n|x)}
  =\prod_{i=1}^{3}T_{n,i},
\end{equation}
where
\[
  T_{n,1}= \frac{1-k\gamma(x)}{1-k\widetilde{\gamma}_n(x)}, \quad
  T_{n,2}=\frac{\frac{(e(\alpha_n|x))^k}{1-k\gamma(x)}}
  {{\RECTM}_k(\alpha_n|x)}, \quad
  T_{n,3} =  \left(\frac{\hat{e}_n(\alpha_n|x)}{e(\alpha_n|x)}\right)^k.
\]

We begin with the term $T_{n,1}$. Define
$N_n=\sqrt{nh_n^p(1-\alpha_n)}\left(\widetilde{\gamma}_n(x)-\gamma(x)\right)$,
then it follows from Corollary \ref{cor2.1} that $N_n=O_{\P}(1)$ and
\begin{equation}\label{eq5.5}
T_{n,1}=1+\frac{k(nh_n^p(1-\alpha_n))^{-1/2}N_n}
{1-k\left((nh_n^p(1-\alpha_n))^{-1/2}N_n+\gamma(x)\right)}.
\end{equation}
For the term $T_{n,2}$, applying Lemma \ref{lem5.4} with $t=e(\alpha_n|x)$ we have
\begin{equation*}
  \lim_{n\rightarrow\infty}{\frac{1}{A(1/\bar{F}(e(\alpha_n|x)|x)|x)}}
  \left\{
  \frac{{\RECTM}_k(\alpha_n|x)}{(e(\alpha_n|x))^k}
  -\frac{1/\gamma(x)}{1/\gamma(x)-k}
  \right\}
  =\frac{k}{(1/\gamma(x)-k)(1/\gamma(x)-k-\rho(x))}.
\end{equation*}
By using $\sqrt{nh_n^p(1-\alpha_n)}A(1/\bar{F}(e(\alpha_n|x)|x)|x)\rightarrow0$
as $n\rightarrow\infty$, we obtain
\begin{equation*}
  \frac{{\RECTM}_k(\alpha_n|x)}
  {\frac{(e(\alpha_n|x))^k}{1-k\gamma(x)}}-1=o((nh_n^p(1-\alpha_n))^{-1/2}),
\end{equation*}
which implies that
\begin{equation}\label{eq5.6}
  T_{n,2}=1+o((nh_n^p(1-\alpha_n))^{-1/2}).
\end{equation}

Now, combining \eqref{eq5.4}-\eqref{eq5.6}, for large $n$ we have
{\scriptsize\begin{eqnarray*}
  && \sqrt{nh_n^p(1-\alpha_n)}
  \left\{
  \frac{\widetilde{\RECTM}_{k,n}(\alpha_n|x)}{\RECTM_k(\alpha_n|x)}-1
  \right\} \\
  &=&\sqrt{nh_n^p(1-\alpha_n)}
  \Bigg\{
  \left(\left(\frac{\hat{e}_n(\alpha_n|x)}{e(\alpha_n|x)}\right)^k-1\right)
  \left(1+\frac{k(nh_n^p(1-\alpha_n))^{-1/2}N_n}
  {1-k\left((nh_n^p(1-\alpha_n))^{-1/2}N_n+\gamma(x)\right)}\right)
  \left(1+o((nh_n^p(1-\alpha_n))^{-1/2})\right)\\
  &&\qquad\qquad\qquad\qquad+
  \left(1+\frac{k(nh_n^p(1-\alpha_n))^{-1/2}N_n}
  {1-k\left((nh_n^p(1-\alpha_n))^{-1/2}N_n+\gamma(x)\right)}\right)
  \left(1+o((nh_n^p(1-\alpha_n))^{-1/2})\right)-1
  \Bigg\}\\
  &=& \sqrt{nh_n^p(1-\alpha_n)}
  \Bigg\{
  \left(\left(\frac{\hat{e}_n(\alpha_n|x)}{e(\alpha_n|x)}\right)^k-1\right)
  \left(1+o_{\P}(1)\right)
  +\frac{k(nh_n^p(1-\alpha_n))^{-1/2}N_n}
  {1-k\left((nh_n^p(1-\alpha_n))^{-1/2}N_n+\gamma(x)\right)}
  +o((nh_n^p(1-\alpha_n))^{-1/2})
  \Bigg\}.
\end{eqnarray*}}
Recall that $N_n=\sqrt{nh_n^p(1-\alpha_n)}\left(\widetilde{\gamma}_n(x)-\gamma(x)\right)=O_{\P}(1)$,
then it suffices to prove the result of the theorem for the sequence of random vectors
\begin{equation*}
  \sqrt{nh_n^p(1-\alpha_n)}
  \left\{
  \left(\left(\frac{\hat{e}_n(\alpha_n|x)}{e(\alpha_n|x)}\right)^k-1
  +\frac{k}{1-k\gamma(x)}\left(\widetilde{\gamma}_n(x)-\gamma(x)\right)\right), \left(\frac{\hat{e}_n(\alpha_n|x)}{e(\alpha_n|x)}-1\right)
  \right\}^T.
\end{equation*}
Let $\beta=(\beta_1,\beta_2)^T\neq0$ in $\R^2$.
By the Taylor expansion and Corollary \ref{th2.1}, for large $n$ we have
{\footnotesize\begin{eqnarray*}
  &&\sqrt{nh_n^p(1-\alpha_n)}
  \left\{
  \beta_1\left(\left(\frac{\hat{e}_n(\alpha_n|x)}{e(\alpha_n|x)}\right)^k-1
  +\frac{k}{1-k\gamma(x)}\left(\widetilde{\gamma}_n(x)-\gamma(x)\right)\right) +\beta_2\left(\frac{\hat{e}_n(\alpha_n|x)}{e(\alpha_n|x)}-1\right)
  \right\} \\
  &=& \sqrt{nh_n^p(1-\alpha_n)}
  \left\{
  \beta_1\left(\left(\frac{\hat{e}_n(\alpha_n|x)}{e(\alpha_n|x)}\right)^k-1\right)
  +\beta_2\left(\frac{\hat{e}_n(\alpha_n|x)}{e(\alpha_n|x)}-1\right)
  \right\}
  +\frac{\beta_1k}{1-k\gamma(x)}\sqrt{nh_n^p(1-\alpha_n)}
  \left(\widetilde{\gamma}_n(x)-\gamma(x)\right) \\
  &=& (\beta_1 k+\beta_2)\sqrt{nh_n^p(1-\alpha_n)}
  \left(\frac{\hat{e}_n(\alpha_n|x)}{e(\alpha_n|x)}-1\right)(1+o_{\P}(1))
  +\frac{\beta_1k}{1-k\gamma(x)}\sqrt{nh_n^p(1-\alpha_n)}
  \left(\widetilde{\gamma}_n(x)-\gamma(x)\right) \\
  &\overset{d}{\longrightarrow}&
  \N\left(0,\frac{\Vert K\Vert_2^2\gamma^2(x)}{g(x)}\theta^T\Lambda(x)\theta\right),
\end{eqnarray*}}
where $\Lambda(x)$ is given by Theorem \ref{th2.1} and
$\theta=\left(\beta_1k+\beta_2,\frac{\beta_1k}{1-k\gamma(x)}\right)^T$.
After elementary computations, we get $\theta^T\Lambda(x)\theta=\beta^TV(x)\beta$. Hence, the
desired result follows from a Cram\'{e}r-Wold device.
\qed

\noindent
{\bf Proof of Theorem \ref{th2.3}.}
Let us write
{\small\begin{equation}\label{eq5.7}
\frac{\sqrt{nh_n^p(1-\alpha_n)}}{\log((1-\alpha_n)/(1-\beta_n))}
\left\{\log\widetilde{\RECTM}_{k,n}^W(\beta_n|x)-\log{\RECTM}_k(\beta_n|x)\right\}
=\sum_{i=1}^3 Q_{n,i},
\end{equation}}
where
{\footnotesize\begin{eqnarray*}
  Q_{n,1} &=& k\sqrt{nh_n^p(1-\alpha_n)}\left(\widetilde{\gamma}_n(x)-\gamma(x)\right), \\
  Q_{n,2} &=& \frac{\sqrt{nh_n^p(1-\alpha_n)}}{\log((1-\alpha_n)/(1-\beta_n))}
  \log\left(\frac{\widetilde{\RECTM}_{k,n}(\alpha_n|x)}{{\RECTM}_k(\alpha_n|x)}\right), \\
  Q_{n,3} &=& \frac{\sqrt{nh_n^p(1-\alpha_n)}}{\log((1-\alpha_n)/(1-\beta_n))}
  \Big(
  \log{\RECTM}_k(\alpha_n|x)-\log{\RECTM}_k(\beta_n|x)
  +k\gamma(x)\log((1-\alpha_n)/(1-\beta_n))
  \Big).
\end{eqnarray*}}
First, according to Corollary \ref{cor2.1}, for large $n$ we have
\begin{equation}\label{eq5.8}
  Q_{n,1}\overset{d}{\longrightarrow}
  \N\left(0,\frac{k^2\Vert K\Vert_2^2\gamma^2(x)}{g(x)}\Lambda_{2,2}(x)\right),
\end{equation}
where $\Lambda_{2,2}(x)$ is given by \eqref{eq2.35}.
Next, from Corollary \ref{cor2.2}, we have
\begin{eqnarray}\label{eq5.9}
  Q_{n,2}&=&\frac{\sqrt{nh_n^p(1-\alpha_n)}}{\log((1-\alpha_n)/(1-\beta_n))}
  \left(\frac{\widetilde{\RECTM}_{k,n}(\alpha_n|x)}{{\RECTM}_k(\alpha_n|x)}-1\right)
  (1+o_{\P}(1))\nonumber\\
  &=&\frac{1}{\log((1-\alpha_n)/(1-\beta_n))}O_{\P}(1)\nonumber\\
  &=&o_{\P}(1),
\end{eqnarray}
where in the last equality we used that $\frac{1-\beta_n}{1-\alpha_n}\rightarrow0$
as $n\rightarrow\infty$. Now, Lemma \ref{lem5.2} implies that
{\footnotesize\begin{eqnarray}\label{eq5.10}
  Q_{n,3}
  &=&\frac{\sqrt{nh_n^p(1-\alpha_n)}}{\log((1-\alpha_n)/(1-\beta_n))}
  \log\left(1+\frac{k\gamma(x)}{e(\alpha_n|x)}
  \big(m^{(1)}(x)+o(1)\big)
  +O\big(A((1-\alpha_n)^{-1}|x)\big)
  +O\big(A(1/\bar{F}(e(\alpha_n|x)|x)|x)\big)
  \right) \nonumber\\
  &=&\frac{k\gamma(x)\sqrt{nh_n^p(1-\alpha_n)}}{e(\alpha_n|x)\log((1-\alpha_n)/(1-\beta_n))}
  \big(m^{(1)}(x)+o(1)\big)
  +O\left(\frac{A((1-\alpha_n)^{-1}|x)\sqrt{nh_n^p(1-\alpha_n)}}
  {\log((1-\alpha_n)/(1-\beta_n))}\right) \nonumber\\
  &&+O\left(\frac{A(1/\bar{F}(e(\alpha_n|x)|x)|x)\sqrt{nh_n^p(1-\alpha_n)}}
  {\log((1-\alpha_n)/(1-\beta_n))}\right)\nonumber\\
  &=&o(1),
\end{eqnarray}}
where the last equality follows by using the assumptions that
$\frac{\sqrt{nh_n^p(1-\alpha_n)}}{e(\alpha_n|x)}\rightarrow\lambda(x)\in\R$,
$\frac{1-\beta_n}{1-\alpha_n}\rightarrow0$,
$\sqrt{nh_n^p(1-\alpha_n)}A((1-\alpha_n)^{-1}|x)\rightarrow0$
and
$\sqrt{nh_n^p(1-\alpha_n)}A(1/\bar{F}(e(\alpha_n|x)|x)|x)\rightarrow0$
as $n\rightarrow\infty$.
Finally, combining \eqref{eq5.7}-\eqref{eq5.10}, the result of
Theorem \ref{th2.3} follows.
\qed

\end{document}